\documentclass[%
 reprint,
 amsmath,amssymb,
 aps,
]{revtex4-2}
\usepackage[utf8]{inputenc}
\usepackage{graphicx}
\usepackage{dcolumn}
\usepackage{bm}
\usepackage{physics}
\usepackage{url}
\usepackage{hyperref}
\hypersetup{
 setpagesize=false,
 bookmarksnumbered=true,%
 bookmarksopen=true,%
 colorlinks=true,%
 linkcolor=blue,
 citecolor=blue,
}

\begin{document}

\preprint{APS/123-QED}

\title{Relativistic resistive magneto-hydrodynamics code for high-energy heavy-ion collisions}

\author{Kouki Nakamura$^{1,2}$}
 \email{knakamura@hken.phys.nagoya-u.ac.jp}
\author{Takahiro Miyoshi$^{2}$}%
 \email{miyoshi@sci.hiroshima-u.ac.jp}
\author{Chiho Nonaka$^{1,2,3}$}%
 \email{nchiho@hiroshima-u.ac.jp}
\author{Hiroyuki R. Takahashi$^{4}$}%
 \email{takhshhr@komazawa-u.ac.jp}
\affiliation{%
$^1$Department of Physics, Nagoya University, Nagoya 464-8602, Japan \\
$^2$Department of Physics, Hiroshima University, Higashihiroshima 739-8526, Japan\\
$^3$Kobayashi Maskawa Institute, Nagoya University, Nagoya 464-8602, Japan\\
$^4$Department of Physics, Komazawa University, Tokyo 154-8525, Japan\\
 }%




\date{\today}

\begin{abstract}
We construct a relativistic resistive magneto-hydrodynamic (RRMHD) numerical simulation code for high-energy heavy-ion collisions. 
We split the system of differential equations into two parts, a non-stiff and a stiff part.
For the non-stiff part, we evaluate the numerical flux using HLL approximated Riemann solver and execute the time integration by the second-order of Runge-Kutta algorithm.
For the stiff part, which appears in Ampere's law, we integrate the equations using semi-analytic solutions of the electric field.
We employ the generalized Lagrange multiplier method to ensure the divergence-free constraint for the magnetic field and Gauss's law.
We confirm that our code reproduces well the results of standard RRMHD tests in the Cartesian coordinates.
In the Milne coordinates, the code with high conductivity is validated against relativistic ideal MHD tests.
We also verify the semi-analytic solutions of the accelerating longitudinal expansion of relativistic resistive magneto-hydrodynamics in high-energy heavy-ion collisions in a comparison with our numerical result. Our numerical code reproduces these solutions.
\end{abstract}

\maketitle
\section{introduction}

Relativistic hydrodynamics has been widely used for description of collective dynamics in various phenomena from nuclear physics to astrophysics.
The high-energy heavy-ion collision experiment is one of the active fields of application of relativistic hydrodynamics.

At Relativistic Heavy Ion Collider (RHIC), 
production of the strongly interacting quark-gluon plasma (QGP) was succeeded, which was achieved by measurement of key observables and theoretical interpretation to them~\cite{BACK200528,ADCOX2005184,ADAMS2005102,ARSENE20051}. 
In particular, at that time, the strong elliptic flow was successfully explained by hydrodynamic models. 
Along with other phenomenological analyses, this fact  reached the conclusion that the QGP created at RHIC is not weakly interacting gas but strongly interacting plasma. 
The analysis of high-energy heavy-ion collisions based on the relativistic viscous hydrodynamics shed light on not only understanding dynamics of space-time evolution after collisions, but also QGP bulk properties such as temperature dependence of shear and bulk viscosities and a diffusion constant.
Usually, the QGP bulk properties are discussed in comparison between experimental data and hydrodynamic model calculation.  
Intensive computation of Bayesian analysis is also performed for extracting detailed information of thermodynamic properties of QCD~\cite{Bernhard:2019bmu,PhysRevC.103.054904}. 

During the relativistic collision of positively charged heavy nuclei, the highest intense electromagnetic fields in our universe are produced, e.g. $|eB|\sim 10^{15} ~\mathrm{T}$~\cite{Huang_2016}.
The high-energy heavy-ion collision is possible to address the property of interaction between deconfined strongly interacting matter and electromagnetic fields. 
The effect of electrical conductivity of the medium on the evolution of electromagnetic fields has been discussed with the semi-analytic solutions of electromagnetic fields in Refs.~\cite{PhysRevC.88.024911,MCLERRAN2014184}.
The contribution of electromagnetic fields to the charge-dependent anisotropic flow has been estimated from the semi-analytic solution of simplified relativistic hydrodynamic equations with electromagnetic fields in Refs.~\cite{Hirono:2012rt,Gursoy:2014aka,Gursoy:2018yai}.
Furthermore, relativistic hydrodynamic simulation with background electromagnetic fields has been investigated~\cite{Pang:2016yuh,Roy:2017yvg}.

The fully realistic analysis of high-energy heavy-ion collisions would require one to solve Maxwell equations together with relativistic hydrodynamic equations.
It is called a full relativistic magneto-hydrodynamic (RMHD) framework, which describes the dynamics of the plasma coupled with electromagnetic fields.
Such an analysis with infinite electrical conductivity, relativistic ideal MHD, has been carried out~\cite{Inghirami:2016iru,Inghirami:2019mkc}.
In the relativistic ideal MHD, the electric field measured in the fluid comoving frame vanishes.
In order to consider the charge distribution of hadrons in high-energy heavy-ion collisions, the simulation based on the RMHD framework with finite electrical conductivity is needed.
Therefore, we construct a model for high-energy heavy-ion collisions based on relativistic resistive magneto-hydrodynamics (RRMHD).

The RRMHD simulation has been performed in astrophysical situations such as black hole accretion disks, jets, and neutron star magnetospheres. 
The dissipation associated with Ohmic conduction current affects the topology of the magnetic field and liberates the magnetic energy, which is known as magnetic reconnection discussed in astrophysical applications~\cite{Takahashi_2011,Zenitani_2010,Watanabe_2006}.

In astrophysical community, there are two known difficulties for constructing the RRMHD code. 
First one is that a time step size $\Delta t$ becomes too much short for calculation of highly conducting medium with explicit time integration.
Then it is time consuming to follow a long time evolution of dynamics. 
Several numerical schemes are proposed to solve this problem, such as the semi-analytic method ~\cite{Komissarov:2007wk} or the semi-implicit scheme~\cite{10.1111/j.1365-2966.2009.14454.x}.
Another difficulty is found in keeping the constraints of the Maxwell equations, e.g., Gauss’s law and the divergence-free constraint.
Ampere’s law and Faraday’s law describe the time evolution of electric and magnetic fields, respectively.
These equations ensure that constraints hold if they are satisfied at the initial state.
These conditions are, however, not preserved numerically in a not well-designed scheme for multi-dimensional problems. 
The violation of the constraints leads to the growth of spurious oscillation and the simulation is unexpectedly terminated. 
In the framework of RRMHD, the generalized Lagrange multiplier (GLM) method is proposed to guarantee these conditions ~\cite{Komissarov:2007wk,MUNZ2000484}. 
In this scheme, the numerical errors relating to the violation of the constraints are advected and dissipated. 
The GLM scheme is commonly used also in MHD~\cite{DEDNER2002645} and RMHD~\cite{Porth:2016rfi} and shows the robustness of the numerical code. 

In this paper, we construct RRMHD code by adopting the semi-analytic method for time integration and GLM method for preserving the conditions in the Milne coordinates for the purpose of understanding high-energy heavy-ion collisions.
We check the correctness and robustness of our code by performing several numerical tests which are common test problems for relativistic ideal MHD and RRMHD. 
We also propose the test problem of longitudinal expansion with an acceleration of relativistic resistive magneto-hydrodynamics. 
The semi-analytic solution of this problem is shown by mimicking the electromagnetic configuration and fluid velocity in high-energy heavy-ion collisions~\cite{PhysRevD.102.014017}. 
We apply this solution as a test problem of RRMHD in the Milne coordinates.

This paper is organized as follows.
In Sec.~\ref{RRMHD}, we briefly review the formulation of RRMHD.
We explain the numerical algorithm of the RRMHD systems in our simulation code in Sec.~\ref{Numerical algorithm}.
We show numerical benchmark tests as a verification of our RRMHD simulation code in Sec.~\ref{test problem}.
A summary is given at the end in Sec.~\ref{summary}.
Unless otherwise specified, we use natural units $\hbar = c = \epsilon_0 = \mu_0 = 1$, where $\epsilon_0$ and $\mu_0$ are the electric permittivity and the magnetic permeability in vacuum, respectively.
Throughout the paper, the components of the four tensors are indicated with greek indices, whereas three vectors are denoted as boldface symbols. 

\section{Relativistic resistive magneto-hydrodynamics}\label{RRMHD}

The conservation laws for the charged current $N^\mu$ and for the total energy-momentum tensor of the plasma $T^{\mu\nu}$ in the dynamics of the whole system are written by,
\begin{eqnarray}
  \label{ccc}
  \nabla_\mu N^\mu = 0,\\
  \label{emc}
  \nabla_\mu T^{\mu\nu} = 0,
\end{eqnarray}
where $\nabla_\mu$ is the covariant derivative.
The electromagnetic fields follow Maxwell equations,
\begin{eqnarray}
  \label{maxwell1}
  \nabla_\mu F^{\mu\nu} = -J^{\nu},\\
  \label{maxwell2}
  \nabla_\mu ~^\star F^{\mu\nu} = 0,
\end{eqnarray}
where $F^{\mu\nu}$ is a Faraday tensor and $^{\star}F^{\mu\nu} = \frac{1}{2}\epsilon^{\mu\nu\rho\sigma}F_{\mu\nu}$ is its dual tensor,
with $\epsilon^{\mu\nu\rho\sigma} = (-g)^{-1/2}[\mu\nu\rho\sigma]$, $g = \det(g_{\mu\nu}) $ and $[\mu\nu\rho\sigma]$ is a completely anti-symmetric tensor. 
Here $g_{\mu\nu}$ is a metric tensor.
If the magnetization and polarization effects are ignored, the energy-momentum tensor of the electromagnetic fields is written by,
\begin{eqnarray}
  T^{\mu\nu}_{f} = F^{\mu\lambda}F_\lambda^\nu - \frac{1}{4}g^{\mu\nu}F^{\lambda\kappa}F_{\lambda\kappa},
\end{eqnarray}
and this tensor follows $\nabla_\mu T^{\mu\nu}_f = J_{\mu}F^{\mu\nu}$, from Maxwell equations.
The total energy-momentum tensor is the sum of the contribution of matter and electromagnetic fields $T^{\mu\nu} = T^{\mu\nu}_m + T^{\mu\nu}_f$.
The conservation law of the total system shown by Eq.~(\ref{emc}) gives,
\begin{eqnarray}
  \nabla_\mu T^{\mu\nu}_m = -J_{\mu}F^{\mu\nu}.
\end{eqnarray}
For the ideal fluid, 
the energy-momentum tensor and the charged current of fluids are written by,
\begin{eqnarray}
  N^\mu = \rho u^\mu,\\
  T^{\mu\nu}_m = (e+p)u^\mu u^\nu +pg^{\mu\nu},
\end{eqnarray}
where $u^{\mu}$ ($u^\mu u_\mu = -1$) is a single fluid four-velocity, $\rho$ is the fluid density, $e = T^{\mu\nu}_m u_\mu u_\nu$ is energy density and $p =\frac{1}{3}\Delta_{\mu\nu}T^{\mu\nu}_m$ is pressure of the fluid.
We have introduced the projection tensor $\Delta_{\mu\nu}= g_{\mu\nu} + u_\mu u_\nu$.
The Faraday tensor and its dual tensor can be rewritten as,
\begin{eqnarray}
        F^{\mu\nu}= u^\mu e^\nu - u^\nu e^\mu + \epsilon^{\mu\nu\lambda\kappa}b_\lambda u_\kappa,\\
  ^\star F^{\mu\nu} = u^{\mu} b^{\nu} - u^{\nu} b^{\mu} - \epsilon^{\mu\nu\lambda\kappa}b_\lambda u_\kappa, 
\end{eqnarray}
where,
\begin{eqnarray}
  e^\mu = F^{\mu\nu}u_\nu,~( e^{\mu}u_\mu = 0 ),\\
  b^\mu = ~^\star F^{\mu\nu}u_\nu,~( b^{\mu}u_\mu = 0),
\end{eqnarray}
are the electric and magnetic fields measured in the comoving frame of the fluid.
We introduce the electric field $E^i$ and the magnetic field $B^i$, which are related to $e^\mu$ and $b^\mu$ by
\begin{eqnarray}
  e^\mu = (\gamma v_kE^k, \gamma E^i + \gamma\epsilon^{ijk}v_jB_k),\\
  b^\mu = (\gamma v_kB^k, \gamma B^i - \gamma\epsilon^{ijk}v_jE_k), 
\end{eqnarray} 
where $\gamma$ is the Lorentz factor of the fluid and $v^i$ is the fluid three-velocity.
The projections of $\nabla_\mu T^{\mu\nu}_m = -J_{\mu}F^{\mu\nu}$ along the perpendicular and parallel directions with respect to $u^{\mu}$ are given by,
\begin{gather}
    (e+p)Du^{\alpha}+(\nabla^\alpha + u^\alpha D)p = g^\alpha_\nu F^{\nu\lambda}J_\lambda - u^{\alpha}e^{\lambda}J_{\lambda},\label{momentum conservation equation}\\
    De + (e+p)\Theta=e^{\lambda}J_{\lambda},\label{energy conservation equation}
\end{gather}
where $D=u^{\mu}\nabla_\mu$ and $\Theta=\nabla_\mu u^{\mu}$.
Equations~(\ref{momentum conservation equation}) and (\ref{energy conservation equation}) correspond to the equation of motion and energy equation, respectively. They are auxiliary equations that are not solved in our numerical code. 
However, they are useful for the discussion of Bjorken flow in resistive medium, shown in Subsections \ref{Bjorken} and \ref{ALE}.

Since the system of equations~Eqs.(\ref{ccc})-(\ref{maxwell2}) is closed by Ohm's law, we adopt the simplest form of it shown in Ref.~\cite{Blackman:1993pbp}. In the covariant form, Ohm's law is written by,
\begin{equation}\label{ohmslaw}
  J^\mu = \sigma F^{\mu\nu}u^{\nu} + qu^\mu,
\end{equation}
where $\sigma$ is electrical conductivity and $q = -J^\mu u_\mu$ is electric charge density of the fluid in the comoving frame.
Maxwell equations lead to the charge conservation law,
\begin{equation}
    \partial_\mu J^\mu = 0.
\end{equation}
When we take the ideal limit ($\sigma \rightarrow \infty$) of Ohm's law, Eq.~(\ref{ohmslaw}) is reduced to,
\begin{equation}
    e^{\mu} = 0.
\end{equation}

\section{Numerical procedure}\label{Numerical algorithm}
We now represent governing equations of the RRMHD in a suitable form for numerical calculation. 
Also, we show a numerical scheme to solve RRMHD equations which are appropriate for studying high-energy heavy-ion collisions.

\subsection{Metric}
We split the space-time into 3 + 1 components by space-like hypersurface defined as the iso-surfaces of a scalar time function $t$ and assume a metric of the form,


\begin{equation}
  ds^2= -dtdt + g_{ij}dx^idx^j.\label{metric}
\end{equation}
Since we consider only the Cartesian and the Milne coordinates in this paper, we take $g^{ij} = 0$ ($i \neq j$).
We will use the Cartesian coordinates ($t,x,y,z$) in Subsections \ref{ST}, \ref{CPAW}, \ref{SSCS}, \ref{Cy} and \ref{RT-Min}.
The Milne coordinates ($\tau,x,y,\eta_s$) will be adopted in Subsections~\ref{RT-Milne}, \ref{Bjorken} and \ref{ALE}, where $\tau := \sqrt{t^2-z^2}$ is the longitudinal proper-time and $\eta_s:= \frac{1}{2}\ln{\frac{t+z}{t-z}}$ is the space rapidity.
We note that in the Cartesian coordinates, the three metric $g_{ij} = \mathrm{diag}\{1,1,1\}$, with $\sqrt{-g} = 1$, whereas in the Milne coordinates, $g_{ij} = \mathrm{diag}\{1,1,\tau^2\}$, with $\sqrt{-g} = \tau$.
In both coordinates, $\partial_j g_{ik} =0~(j = 1,2,3)$, source terms of space-components in conservative equations vanish.
However, in the Milne coordinates, where $g_{33} = \tau^2$, the source terms for the energy conservative equation contain a non-zero term proportional to $\frac{1}{2}\partial_0 g_{33} = \tau$. 

\subsection{Constraint equations}
Maxwell equations contain two constraint equations,
\begin{gather}
    \partial_i(\sqrt{-g}E^i) = \sqrt{-g}q,\\
    \partial_i(\sqrt{-g}B^i) = 0.
\end{gather}
Though Maxwell equations ensure that these constraints are satisfied at all time steps, a simple integration of Maxwell equations in numerical simulation does not preserve these conditions because of the numerical error.
It leads to unphysical oscillation and numerical simulation crashes in the end for multi-dimensional problems.
For this reason, a number of numerical techniques for avoiding this problem are investigated.
In this model, we adopt the GLM method to guarantee these conditions~\cite{MUNZ2000484,Komissarov:2007wk,Porth:2016rfi}.
The main idea is that one introduces two variables, $\psi$ and $\phi$ as the deviation from constraints.
One manages a system of equations to decay or carry the deviation $\psi$ and $\phi$ out of the computation domain by relatively high-speed waves.

For GLM method, we extend the Eqs.~(\ref{maxwell1}) and (\ref{maxwell2}) to,
\begin{gather}
    \nabla_\nu(F^{\mu\nu} + g^{\mu\nu}\psi) = -\kappa n^{\mu}\psi + J^{\mu},\label{Mmaxwell1}\\
    \nabla_\nu(^{\star}F^{\mu\nu} + g^{\mu\nu}\phi) = -\kappa n^{\mu}\phi,\label{Mmaxwell2}
\end{gather}
where $\psi$ and $\phi$ are new variables and $\kappa$ is a positive constant.
In the Cartesian coordinates, we obtain the telegraph equations for $\div{\bm{E}}-q$ and $\div{\bm{B}}$,
\begin{gather}
    \partial_t^2(\div{\bm{E}} - q) + \kappa\partial_t(\div{\bm{E}} - q) - \partial_i\partial^i(\div{\bm{E}} - q) = 0,\\
    \partial_t^2(\div{\bm{B}}) + \kappa\partial_t(\div{\bm{B}}) - \partial_i\partial^i(\div{\bm{B}}) = 0.
\end{gather}
Consequently, $\div{\bm{E}}-q$ and $\div{\bm{B}}$ propagate at the speed of light and decay exponentially over a timescale $1/\kappa$.
In the form of the metric Eq.~(\ref{metric}), a time-like normal vector is reduced to $n_\mu = (-1,\bm{0})$.
The modified divergence-free equation of $\bm{B}$ and the Faraday law become,
\begin{gather}
    \partial_t(\sqrt{-g}\phi)+\partial_i(\sqrt{-g}B^i) = -\sqrt{-g}\kappa\phi,\\
    \partial_t(\sqrt{-g}B^j) + \partial_i[\sqrt{-g}(\epsilon^{ijk}E_k + g^{ij}\phi)] = 0.
\end{gather}
Also, the modified Gauss's law and Ampere's law are obtained as,
\begin{gather}
     \partial_t(\sqrt{-g}\psi) + [\partial_i(\sqrt{-g}E^i)-\sqrt{-g}q] = -\sqrt{-g}\kappa\psi,\\
     \partial_t(\sqrt{-g}E^j) - \partial_i[\sqrt{-g}(\epsilon^{ijk}B_k + g^{ij}\psi)] = - \sqrt{-g}J^j.
\end{gather}
The conservation law of electric charge is written by,
\begin{equation}
    \partial_t (\sqrt{-g} q) +\partial_i (\sqrt{-g}J^i) = 0.\label{charge conservation law}
\end{equation}
\subsection{Basic equations}

Let us rewrite the equations of motion Eqs.~(\ref{ccc})-(\ref{emc}), (\ref{Mmaxwell1})-(\ref{Mmaxwell2}) and (\ref{charge conservation law}) in a conservative form which is appropriate for numerical integration,
\begin{equation}\label{conservative form}
  \partial_0 \left(\sqrt{-g}~\bm{U}\right) + \partial_i \left(\sqrt{-g}~\bm{F}^i\right) = \sqrt{-g}~(\bm{S}_e+\bm{S}_s),
\end{equation}
where $\bm{U}, \bm{F}^i$, $\bm{S}_e$ and $\bm{S}_s$ are the sets of conservative variables, fluxes, source terms which are explicitly solved, and source term of the stiff part, respectively.
These variables contain the following components,
\begin{equation}
  \bm{U} = \mqty(D \\\Pi_j\\\varepsilon\\B^j\\E^j\\q\\\psi\\\phi),~
  \bm{F}^i = \mqty(Dv^i\\T^i_j\\\Pi^i\\\epsilon^{jik}E_k + g^{ij}\psi\\-\epsilon^{jik}B_k + g^{ij}\phi\\J^i\\E^i\\B^i)\nonumber,
\end{equation}
\begin{equation}
  \bm{S_e} = \mqty(0\\\frac{1}{2}T^{ik}\partial_j g_{ik}\\-\frac{1}{2}T^{ik}\partial_0g_{ik}\\0\\-qv^i\\0\\0\\0),~
   \bm{S_s} = \mqty(0\\0\\0\\0\\-J^i_c\\0\\-\kappa\psi\\-\kappa\phi),
\end{equation}
where the total momentum $\Pi^i$, the stress tensor $T_{ij}$ and the total energy density $\varepsilon$ are given by,
\begin{eqnarray}
    D &=& \gamma\rho,\\
  \Pi_i &=& (e+p)\gamma^2v_i + \epsilon_{ijk}E^jB^k,\\
  T_{ij} &=& (e+p)\gamma^2v_iv_j + (p+p_{\rm{em}})g_{ij} - E_iE_j - B_iB_j,\nonumber\\
  \\
  \varepsilon &=& (e+p)\gamma^2 - p + p_{\rm{em}},\\
  J_c^i &=& \sigma e^i, 
\end{eqnarray} 
where the electromagnetic energy density $p_{\mathrm{em}}$ is defined as $p_{\rm{em}} = \frac{1}{2}(E^2+B^2)$.
We adopt the operator splitting method for time integration.
Then Eq.~(\ref{conservative form}) can be divided into two equations as,
\begin{gather}
    \partial_0 \left(\sqrt{-g}~\bm{U}\right) + \partial_i \left(\sqrt{-g}~\bm{F}^i\right) = \sqrt{-g}\bm{S}_e,\label{non-stiff parts}\\
   \partial_0 \left(\sqrt{-g}~\bm{U}\right)= \bm{S}_s.\label{stiff parts}
\end{gather}
Equation~(\ref{non-stiff parts}) can be integrated in time with an explicit manner, while Eq.~(\ref{stiff parts}) becomes stiff for large $\sigma$ and $\kappa$.
The one-dimensional discretization of Eq.~(\ref{non-stiff parts}) can be written by,
\begin{equation}
    \bm{U}^{n+1}_i = \bm{U}^{n}_i - \frac{\Delta t}{\Delta x} (\bm{f}_{i + 1/2} - \bm{f}_{i - 1/2}) + \bm{S}_e\Delta t,
\end{equation}
where $\Delta x$ is the grid spacing and $\bm{f}$ is the numerical flux.
The subscript $i$ denotes the grid point, $x = i\Delta x$, and the superscript $n$ shows the number of time steps, $t=n\Delta t$.

We need to reconstruct the primitive variables on the cell surface from those of the cell center to evaluate the numerical flux.
We adopt the second-order accurate scheme for this reconstruction~\cite{VANLEER1977276} given by,
\begin{gather}
    \bm{P}^n_{i \pm 1/2} = \bm{P}^n_{i} \pm \delta \bm{P}^n_i/2,\label{reconstruction}
\end{gather}
\begin{gather}
    \delta \bm{P}^n_i = \left\{
    \begin{array}{l}
    \mathrm{sign}(\delta \bm{P}_{i + 1/2})\min (|\delta \bm{P}_{i+1/2}|/2,2|\delta \bm{P}_{i+1}|,2|\delta \bm{P}_{i}|)\\
    ~~~~~~\mathrm{if}~~\mathrm{sign}(\delta\bm{P}_{i+1})\mathrm{sign}(\delta\bm{P}_{i})>0,\\
    0~~~~\mathrm{otherwise},
    \end{array}
    \right.\label{MC limiter}
\end{gather}
where $\delta \bm{P}_{i+1/2} = \bm{P}_{i+1} - \bm{P}_{i-1}$, $\delta \bm{P}_{i+1} = \bm{P}_{i+1} - \bm{P}_{i}$, and $\delta \bm{P}_{i} = \bm{P}_i - \bm{P}_{i-1}$.
Then the numerical flux is computed by using the Harten-Lax-van Leer (HLL) method~\cite{doi:10.1137/1025002} given by,
\begin{equation}
    \bm{f}_{i \pm 1/2} = \frac{\lambda^+ \bm{F}^{\mathrm{L}}_{i \pm 1/2} - \lambda^- \bm{F}^{\mathrm{R}}_{i \pm 1/2} + \lambda^+\lambda^-(\bm{U}^R_{i \pm 1/2} - \bm{U}^L_{i \pm 1/2})}{\lambda^+ - \lambda^-},\label{HLL flux}
\end{equation}
where $\lambda^+$ and $\lambda^-$ represent the maximum and minimum wave speeds of the system, respectively.
We evaluate them by the speed of light for simplicity.
The superscript $\mathrm{L}$ ($\mathrm{R}$) denotes that the variables are calculated from the primitive variables reconstructed by these at the left (right) grid. 
Then, we numerically integrate Eq.~(\ref{non-stiff parts}) using numerical fluxes.
\subsection{Stiff part}\label{stiff part}
In this subsection, we show how to solve the stiff part Eq.~(\ref{stiff parts}), which contains Ampere's law and equations of $\phi$ and $\psi$.
For Ampere's law, we split the stiff relaxation equation Eq.~(\ref{stiff parts}) into the perpendicular and the parallel directions with respect to $v^i$ and we redefine $E^{'i} = \sqrt{-g}E^i$ and $B^{'i} = \sqrt{-g}B^i$ to,
\begin{gather}
    \partial_0(E_{\parallel}^{'i}) = -\sigma\gamma[E_\parallel^{'i} - (E^{'k} v_k)v^i],\label{parallel}\\
    \partial_0(E_{\perp}^{'i}) = -\sigma\gamma(E_\perp^{'i} + \epsilon^{ijk}v_j B^{'}_k),\label{parp}
\end{gather}
where $E^{'i}_{\perp}$ and $E^{'i}_{\parallel}$ are the perpendicular and the parallel directions of the electric field with respect to $v^i$.
Here, $v^i$ and $\gamma$ are assumed to be a constant during a small time step $\Delta t$.
The solutions of the initial value problem for these relaxation equations are given by,
\begin{gather}
    E_{\parallel}^{'i} = E_{0\parallel}^{'i}\exp{-\sigma\gamma t},\label{parallel solution}\\
    E_{\perp}^{'i} = E_{\perp}^{*i} + (E^{'i}_{0\perp} - E_{\perp}^{*i}) \exp(-\sigma t/\gamma)\label{perp solution},
\end{gather}
where $E^{*i}_{\perp} = -\epsilon^{ijk}v_j B^{'}_k$ and suffix 0 represents the initial value of $E^{'i}$.

For $\psi$ and $\phi$, we can integrate Eq.~(\ref{stiff parts}) as
\begin{eqnarray}
    \psi' = \psi'_0\exp(-\kappa t),\\
    \phi' = \phi'_0\exp(-\kappa t),
\end{eqnarray}
where $\psi' = \sqrt{-g}\psi$ and $\phi'=\sqrt{-g}\phi$, and the suffix 0 represents the initial value of $\psi'$ and $\phi'$.

\subsection{Primitive recovery}\label{Primitive recovery}

In our numerical code, we should reconstruct primitive variables $\{\rho,p,v^i\}$ from evolved conservative variables $\{D,\varepsilon,\Pi_i\}$ in order to calculate numerical flux at each time step.
To reconstruct the primitive variables, we introduce new variables,
\begin{gather}
    \varepsilon' = \varepsilon - p_{em},\\
    \Pi'_i        = \Pi_i - \epsilon_{ijk}E^jB^k,
\end{gather}
where $\varepsilon'$ and $\Pi'$ are the energy density and the momentum of fluid, respectively.
The set of the variables $\{D,\varepsilon',\Pi'_i\}$ is the relativistic ideal fluid conservative variables.
Therefore, we adopt the ordinary primitive reconstruction method of relativistic hydrodynamic simulation for these variables~\cite{AKAMATSU201434}.
We obtain a one-dimensional equation about the gas pressure,
\begin{gather}
    f(p):= [e(p,\rho) +p]\gamma^2(p) - \varepsilon' - p = 0\label{primitive recovery},
    \end{gather}
where,
\begin{gather}
    \frac{1}{\gamma^2(p)} = 1- \frac{\Pi'^i\Pi'_i}{(\varepsilon'+p)^2},\rho = D/\gamma(p).
\end{gather}
The $p$ and $\rho$ dependence of $e$ is determined by the equation of state (EoS). 
This equation is numerically solved by the Newton-Raphson algorithm.
The other primitive variables are reconstructed as,
\begin{gather}
    v^i = \frac{\Pi'^i}{\varepsilon' + p},\\
    \gamma = \frac{1}{\sqrt{1-v^iv_i}},\\
    \rho = D/\gamma,\\
    e = (\varepsilon' +p)/\gamma^2 - p.\label{primitive variable}
\end{gather}

We consider the ideal gas EoS $p = (\Gamma-1)(e-\rho)$ in this paper.
We note that we take the non-zero $\rho$ in Subsections~\ref{ST}, \ref{CPAW}, \ref{SSCS}, \ref{Cy}, and \ref{RT-Min}.
In Subsections~\ref{RT-Milne}, \ref{Bjorken}, and \ref{ALE}, we set $\rho=0$.

\subsection{Numerical algorithm}
Our numerical simulation code is based on finite difference schemes.
The advection equation Eq.~(\ref{non-stiff parts}) including the source term $\bm{S}_e$ is solved by explicit time integration with the second-order TVD Runge-Kutta algorithm~\cite{SHU1988439}.
The primitive variables are interpolated from the cell center to the cell surface by using the second-order accurate scheme represented by Eqs.~(\ref{reconstruction}) and (\ref{MC limiter})~\cite{VANLEER1977276}.
The numerical flux is calculated by the HLL flux shown in Eq.~(\ref{HLL flux})~\cite{doi:10.1137/1025002}.
The stiff part Eq.~(\ref{stiff parts}) is integrated by the analytic solutions explained in Subsection~\ref{stiff part}. 
We adopt the GLM method to guarantee these conditions~\cite{MUNZ2000484,Komissarov:2007wk,Porth:2016rfi} as the divergence cleaning.
The primitive recovery is executed by solving Eq.~(\ref{primitive recovery}) by the Newton-Raphson algorithm explained in Subsection~\ref{Primitive recovery}~\cite{AKAMATSU201434}.
We summarize our numerical algorithm as follows:
\begin{itemize}
    \item[-] the values of the primitive variables are reconstructed from cell center to cell surfaces by using the second-order accurate scheme~\cite{VANLEER1977276} shown in Eqs.~(\ref{reconstruction}) and (\ref{MC limiter}).
    \item[-] conservative variables $\bm{U}$ and fluxes $\bm{F}$ on the cell surface in Eq.~(\ref{conservative form}) are computed.
    \item[-] the Riemann problem for numerical fluxes at cell surfaces is solved by using the HLL flux~\cite{doi:10.1137/1025002} described by Eq.~(\ref{HLL flux}).
    \item[-] Equation~(\ref{non-stiff parts}) is explicitly integrated by using numerical flux.
    \item[-] the stiff equation Eq.~(\ref{stiff parts}) is integrated by using the analytic solutions given by Eqs.~(\ref{parallel solution}) and (\ref{perp solution}).
    \item[-] the new primitive variables are recovered from the evolved conservative variables by solving Eqs.~(\ref{primitive recovery})-(\ref{primitive variable}).
\end{itemize}

\section{test problems}\label{test problem}
In this section, numerical tests are performed for the verification of our numerical simulation code.
We will use the Cartesian coordinates ($t,x,y,z$) in Subsections~\ref{ST}, \ref{CPAW}, \ref{SSCS}, \ref{Cy} and \ref{RT-Min}.
The Milne coordinates ($\tau,x,y,\eta_s$) will be adopted in Subsections~\ref{RT-Milne}, \ref{Bjorken} and \ref{ALE}.
We take the CFL constant as $C_{\mathrm{CFL}} = 0.1$ for all test problems.
\subsection{Shock tube test problem}\label{ST}
In order to test the shock-capturing properties of our numerical simulation code, we consider the simple MHD version of the Brio-Wu test in Ref.~\cite{10.1111/j.1365-2966.2009.14454.x}.
The initial left and right states are given by,
\begin{gather}
    (\rho^L,p^L,(B^y)^L) = (1.0,1.0,1.0)~\mathrm{for}~ x < 0.5,\\
    (\rho^R,p^R,(B^y)^R) = (0.125,0.1,-1.0)~\mathrm{for}~ x \geq 0.5,
\end{gather}
and all the other variables are initially set to 0.

\begin{figure}[ht]
\includegraphics[width=8.5cm,height=6cm]{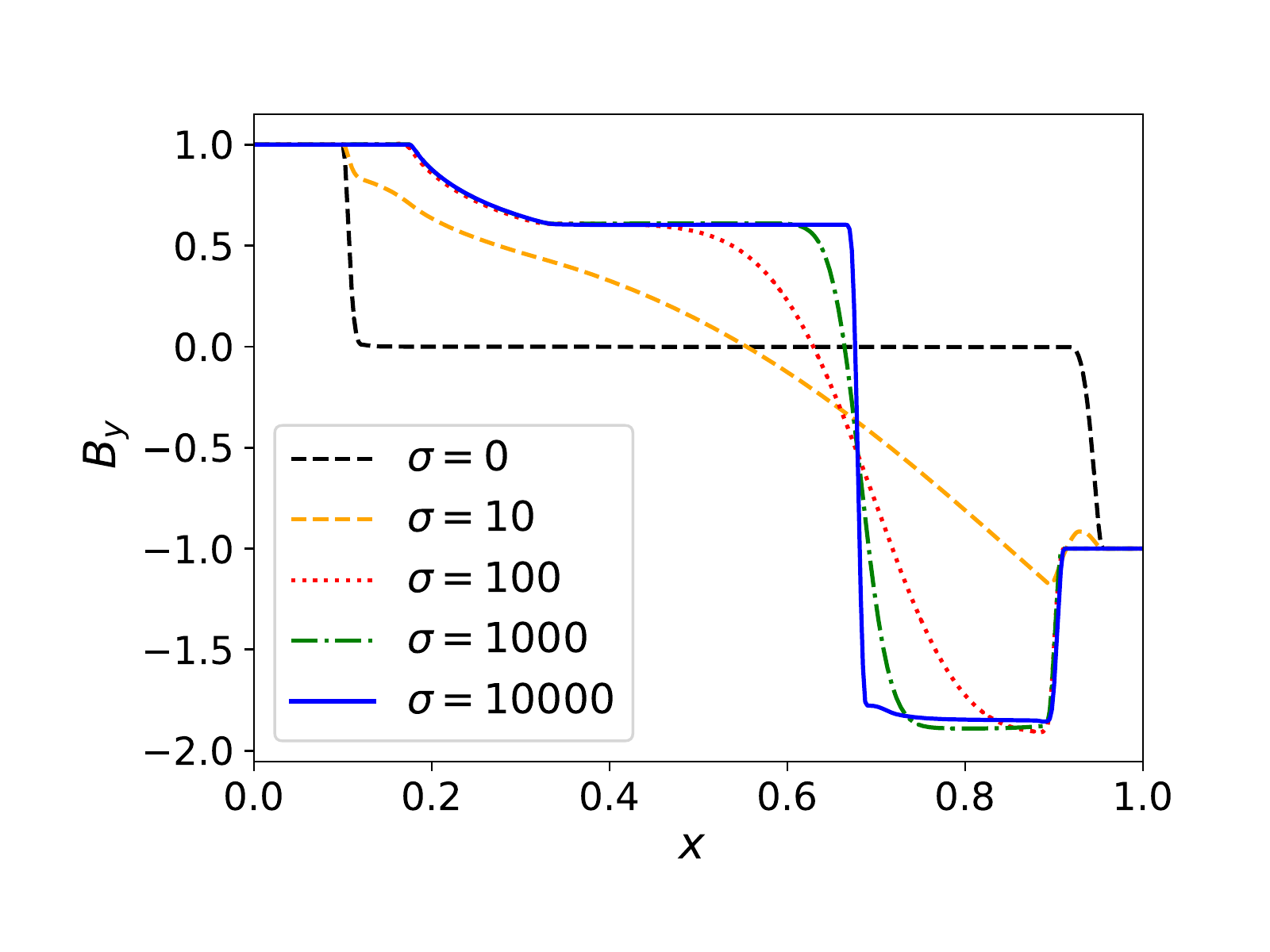}
\caption{\label{fig:Brio-Wu} (color online) We display the magnetic field $B_y$ at $t = 0.4$ in the Brio-Wu type shock tube test problem.}
\end{figure}

Figure~\ref{fig:Brio-Wu} shows the results at $t = 0.4$ for different electrical conductivity, $\sigma = 0,10,10^2,10^3$, and $10^4$.
The simulation box is bounded by $x\in[0, 1.0]$ and the number of grid points is 400.
For $\sigma = 10^4$, the left going rarefaction and right going shock wave, and a tangential discontinuity between them appear.
Our numerical solution with $\sigma = 10^4$ reproduces the solution of relativistic ideal MHD simulations~\cite{giacomazzo_rezzolla_2006}.
In addition, the solutions with low conductivity are similar to the results of the other RRMHD numerical simulations~\cite{10.1111/j.1365-2966.2009.14454.x, Takamoto_2011}.
Especially the electromagnetic field does not interact with gas for $\sigma=0$, so that the electromagnetic waves propagate with light speed.
The wave fronts should be located at  0.1 and 0.9 for left and right going waves, respectively. Our results are consistent with these analytic solutions, although wave fronts slightly have a smooth profile due to the numerical diffusion.

\subsection{Large amplitude circularly polarized Alfv\'{e}n wave}\label{CPAW}
\begin{figure}[ht]
\includegraphics[width=8.5cm,height=6cm]{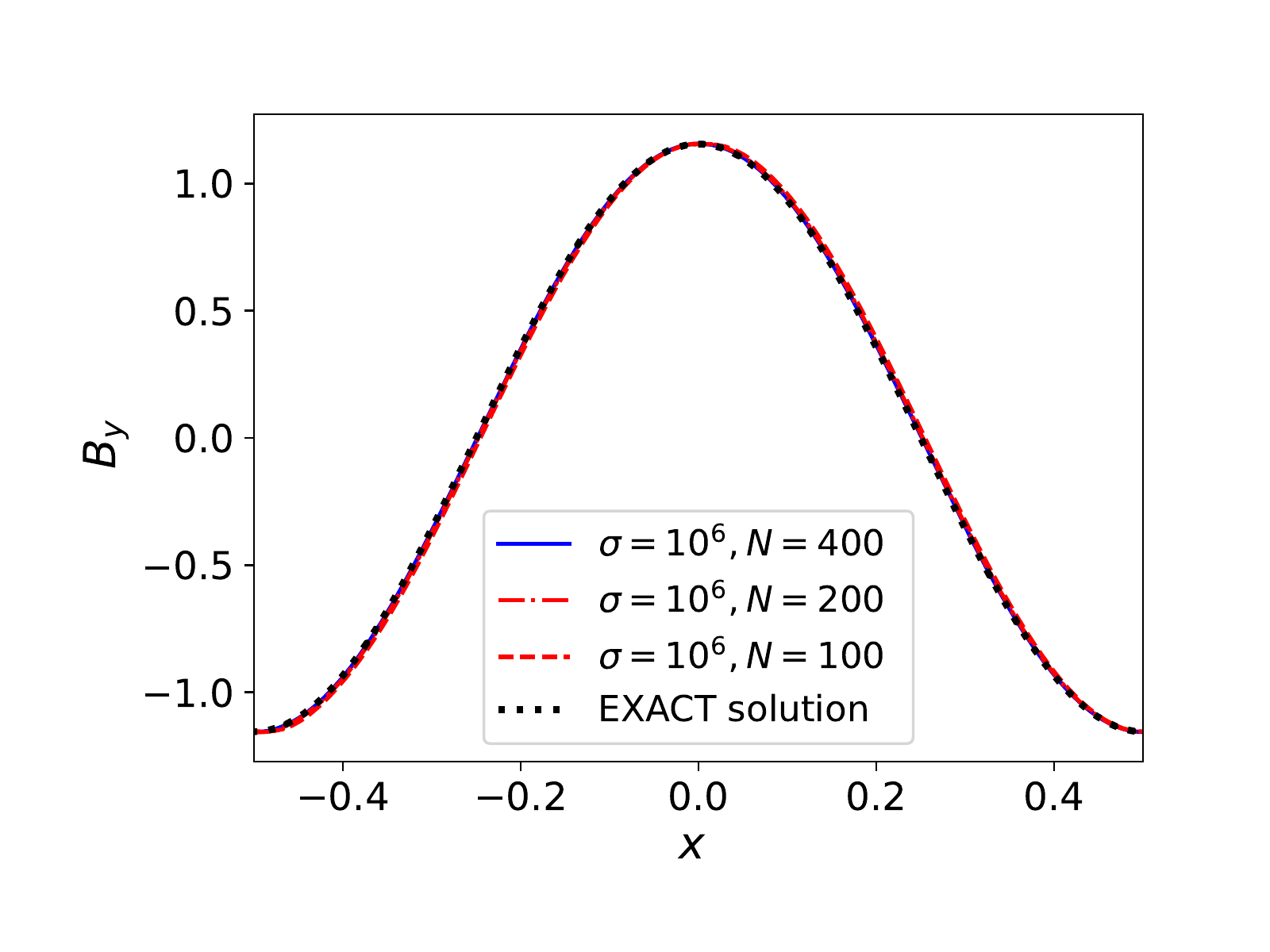}
\caption{\label{fig:LACAW} (color online) The magnetic field component $B_y$ at $t = 2$ is shown in the large amplitude circularly polarized Alf\'{v}en waves. The blue solid, red long dashed-dotted and red dashed lines represent results with $\sigma = 10^6$ and $N=\{100,200,400\}$, respectively. The black dotted line stands for the analytical solution.}
\end{figure}
\begin{figure}[ht]
\includegraphics[width=8.5cm,height=6cm]{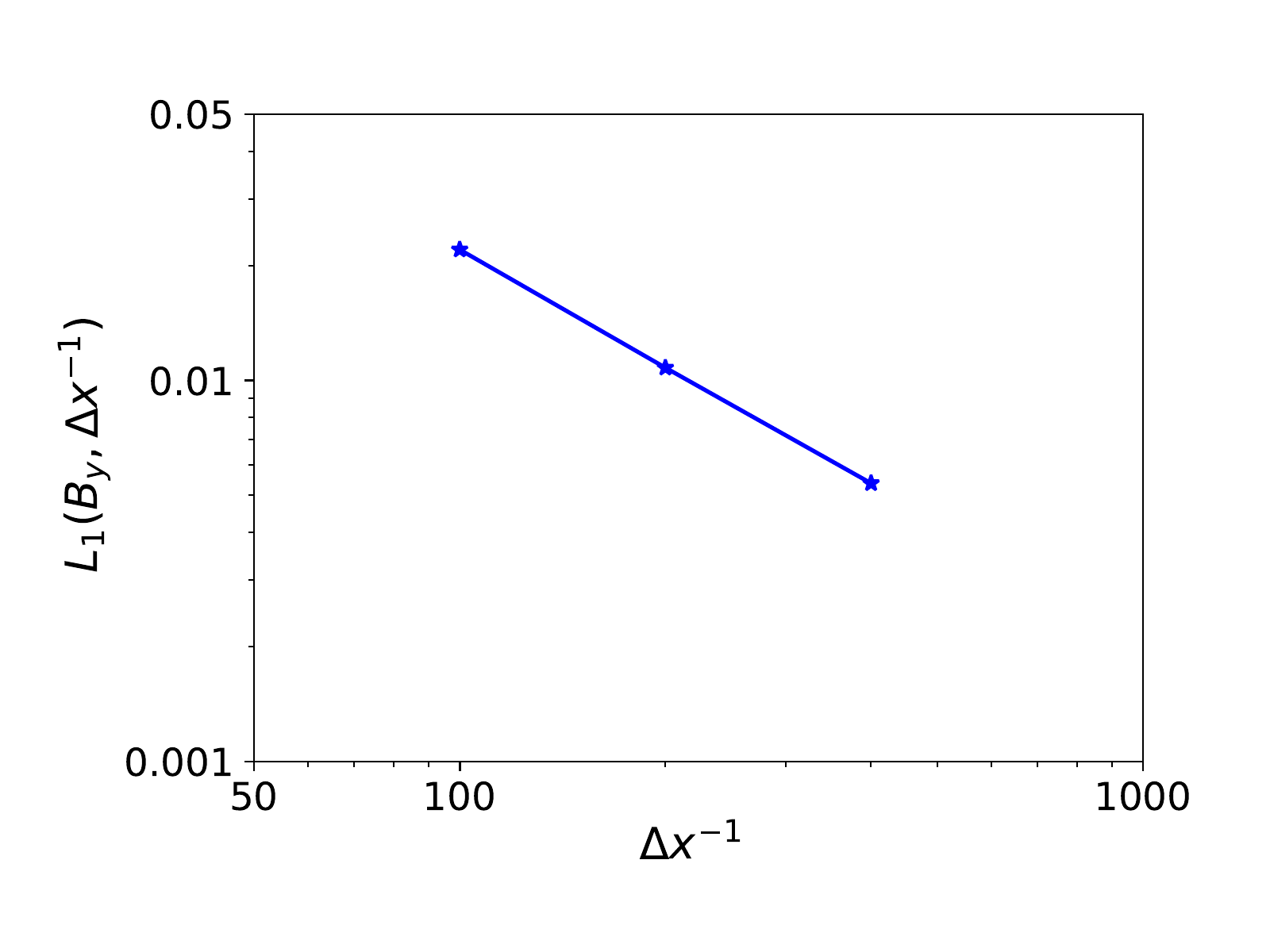}
\caption{\label{fig:LACAW_L1} The $L_1$ norm errors as a function of the inverse of grid-cell size are shown in the large amplitude circularly polarized Alf\'{v}en waves.}
\end{figure}

We perform a test for the propagation of a large amplitude circularly polarized Alf\'{v}en waves along a uniform background magnetic field $B_0$.
The analytical solution of relativistic ideal MHD is proposed in Ref.~\cite{10.1111/j.1365-2966.2009.14454.x}, which is given by,
\begin{gather}
    (B^y,B^z) = \eta_A B^0(\cos[k(x-v_At)],\sin[k(x-v_At)]),\label{CPAW-B}\\
    (v^y,v^z) = -\frac{v_A}{B_0}(B^y,B^z)\label{CPAW-v},
\end{gather}
where $B^x = B_0$, $v_x = 0$, $k = 2\pi$ is the wave number, and $\eta_A$ is the amplitude of the wave.
The square of the special relativistic Alf\'{v}en wave's speed $v^2_A$ is given by,
\begin{equation}
    v_A^2 = \frac{2B_0^2}{h+B_0^2(1+\eta_A^2)}\left[1+\sqrt{1-\left(\frac{2\eta_AB_0^2}{h+B_0^2(1+\eta_A^2)}\right)}\right]^{-1},
\end{equation}
where $h = (e + p)/\rho$ is the specific enthalpy.
We set the initial parameters, $\rho=p=\eta_A = 1$, and $B_0 = 1.1547$.
We use an ideal gas EoS $p=(\Gamma - 1)(e - \rho)$ with $\Gamma = 2$ in this test.
From these parameters, the Alf\'{v}en wave's speed is estimated as $v_A = 1/2$.
The computational domain is $x \in [-0.5,0.5]$.
We use a periodic boundary condition.
Since the solution given by Eqs.~(\ref{CPAW-B}) and (\ref{CPAW-v}) is obtained by solving special relativistic ideal MHD equations, we take a sufficiently large conductivity $\sigma=10^6$ in this test problem.  

Figure~\ref{fig:LACAW} shows our results at $t = 2.0$ (one Alf\'{v}en wave crossing time) for three different number of grid points $N=\{100,200,400\}$ with the analytical solution, in the case of $\sigma = 10^6$.
The blue solid, red long dashed-dotted, and red dashed lines represent results with $N=\{100,200,400\}$, respectively.
The black dotted line stands for the analytical solution.
The results with high conductivity $\sigma$ mean that our simulation code can handle the ideal MHD limit.
Also, there is no dependence of the number of grid points, which suggests that even 100 grid points are enough for the description of the analytical solution. 

For this problem, we cannot achieve full second-order accuracy.
Figure~\ref{fig:LACAW_L1} shows the $L_1(u,\Delta x^{-1})$ norm errors of the tangential magnetic field $B_y$ as a function of the inverse of grid-cell size $\Delta x^{-1}$,
\begin{equation}
    L_1(u,\Delta x^{-1}) := \sum_i^N|u(x_i;N) - u_{\mathrm{EXACT}}(x_i)|\Delta x, 
\end{equation}
where $u_{\mathrm{EXACT}}(x_i)$ is an exact solution at $x = x_i$.
The behavior of $L_1$ norm shows that our numerical simulation is nearly 1.3-order convergence.
The main reason for it is that we execute the second-order Runge-Kutta algorithm with many operator splittings.
It makes the time accuracy of our scheme worsen~\cite{Takamoto_2011}.
This is the most difficulty to solve in RRMHD since this is the limit of large electrical conductivity.

\subsection{Self-similar current sheet}\label{SSCS}
\begin{figure}[ht]
\includegraphics[width=8.5cm,height=6cm]{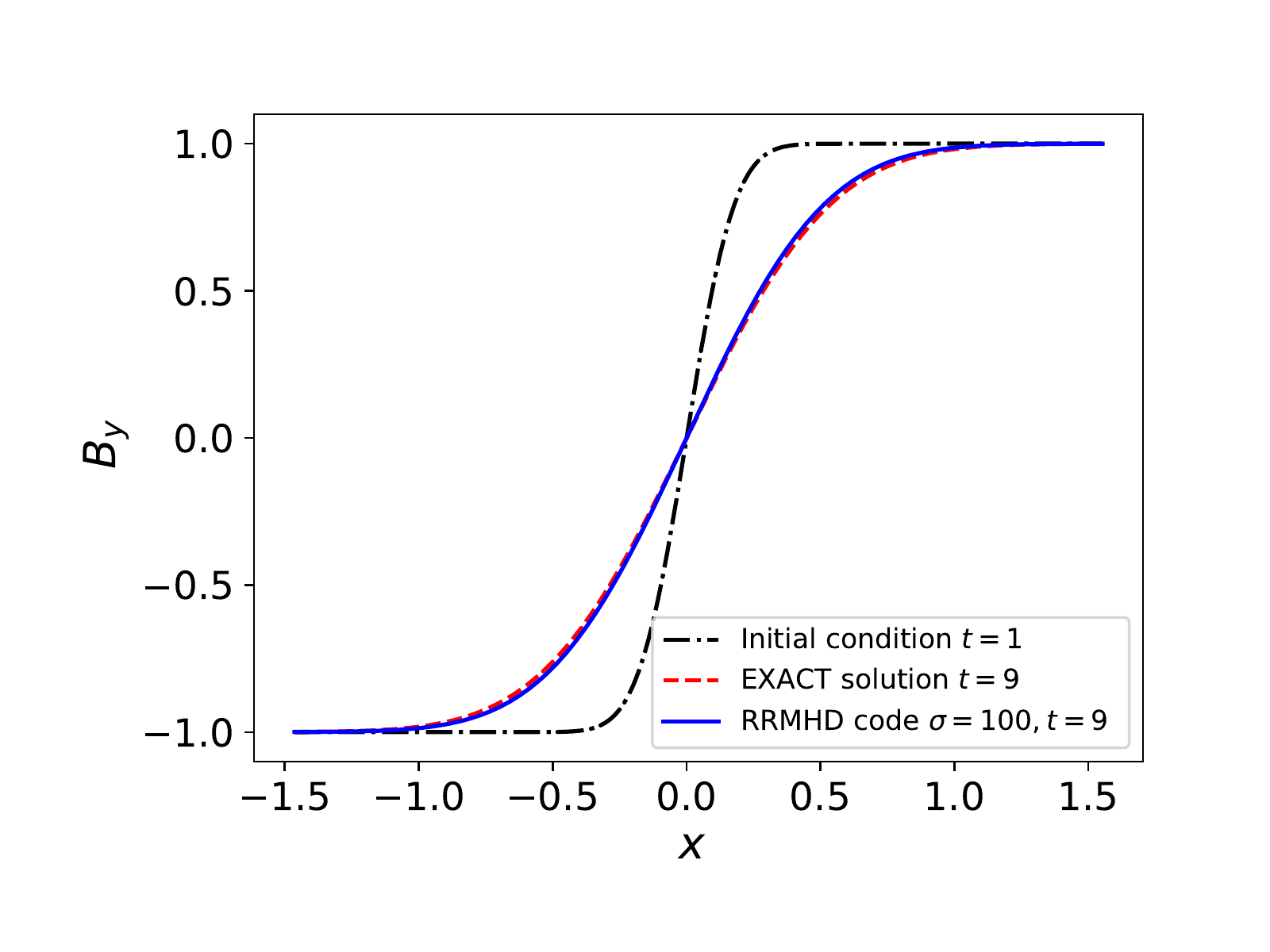}
\caption{\label{fig:SSCS} (color online) The magnetic field component $B_y$ at $t = 9$ in the self-similar current sheet test, with the initial condition at $t=1$.}
\end{figure}

Next, we consider the evolution of the self-similar current sheet proposed in Refs.~\cite{Komissarov:2007wk,10.1111/j.1365-2966.2009.14454.x}.
In this test problem, we assume that the magnetic pressure is much smaller than the pressure of fluid.
The magnetic field has only a tangential component $\bm{B} = (0, B(x,t), 0)$ and $B(x,t)$ changes the sign across the current sheet.
The initial pressure of background fluid is set to be uniform, $p = const$.
We take the high conductivity $\sigma$, so that the diffusion timescale is much longer than the light crossing timescale.
In this assumption, the evolution equation of the magnetic field is reduced to,
\begin{equation}
    \partial_t B -\frac{1}{\sigma}\partial_x^2 B =0.
\end{equation}
The analytic solution of this equation is given by,
\begin{gather}
    B(x,t) = B_0\mathrm{erf}\left(\frac{1}{2}\sqrt{\frac{\sigma x^2}{t}}\right),\\
\end{gather}
where erf is the error function.
We set the initial condition at $t = 1$ with $p = 50$, $\rho = 1$, $\bm{E} = \bm{v} = 0$, and $\sigma = 100$.
The computational domain is $x \in [-1.5,1.5]$, and the number of grid points is $N = 200$.
Figure~\ref{fig:SSCS} shows the numerical result at $t = 9$.
The black long dashed-dotted line represents the initial profile of $B_y$ at $t = 1$.
The blue solid and red dotted lines stand for the numerical result of $B_y$ and the analytical solution at $t = 9$, respectively.
This indicates that the result of our simulation code is consistent with the analytical solution and captures the diffusion of the magnetic field.

\subsection{Cylindrical explosion}\label{Cy}
\begin{figure}[ht]
\includegraphics[width=8.5cm,height=6.5cm]{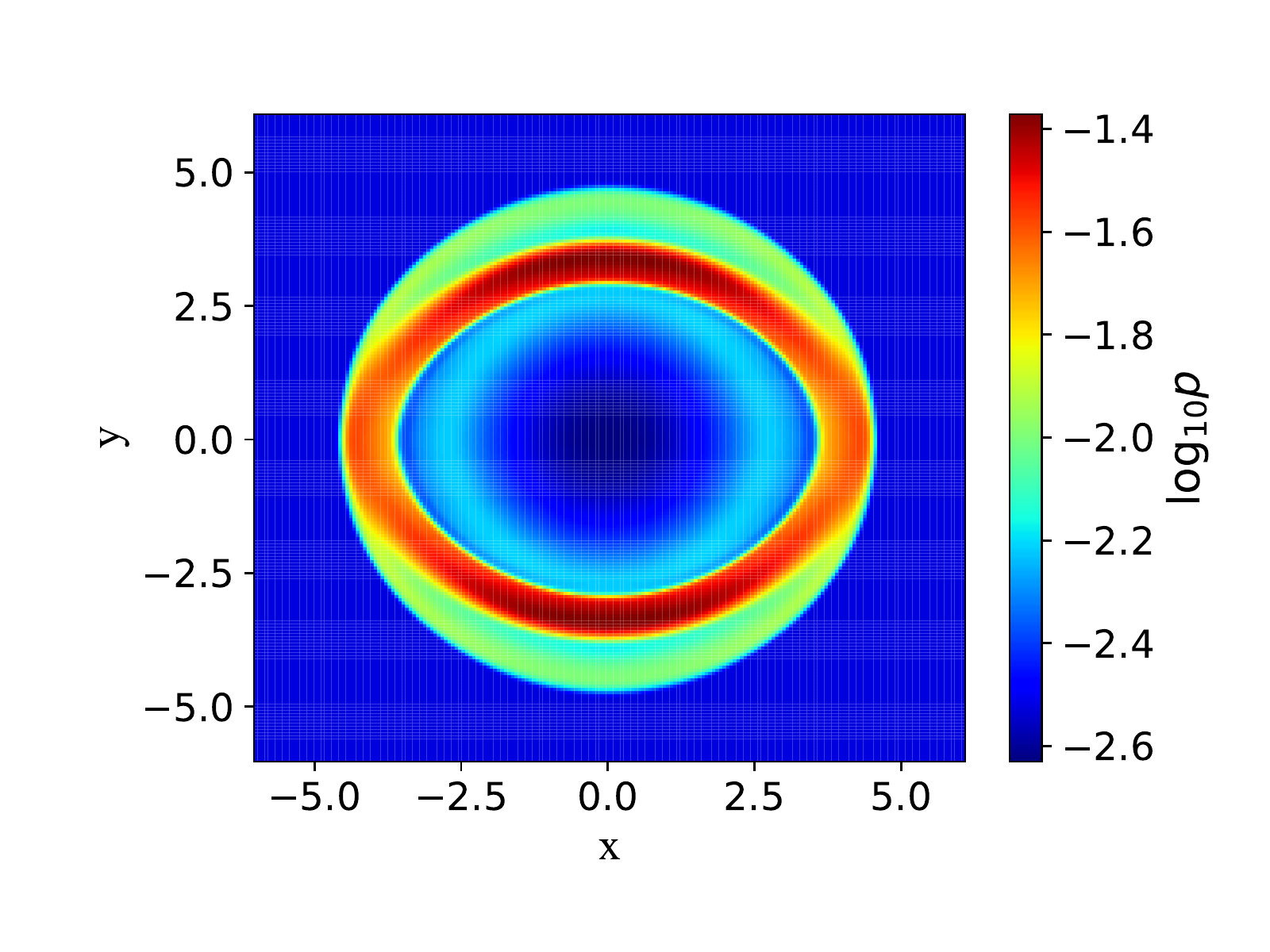}
\caption{\label{fig:CYE_P} The pressure of the fluid at $t = 4$ is shown in the 2D cylindrical explosion problem.}
\end{figure}

\begin{figure}[ht]
\includegraphics[width=8.5cm,height=6.5cm]{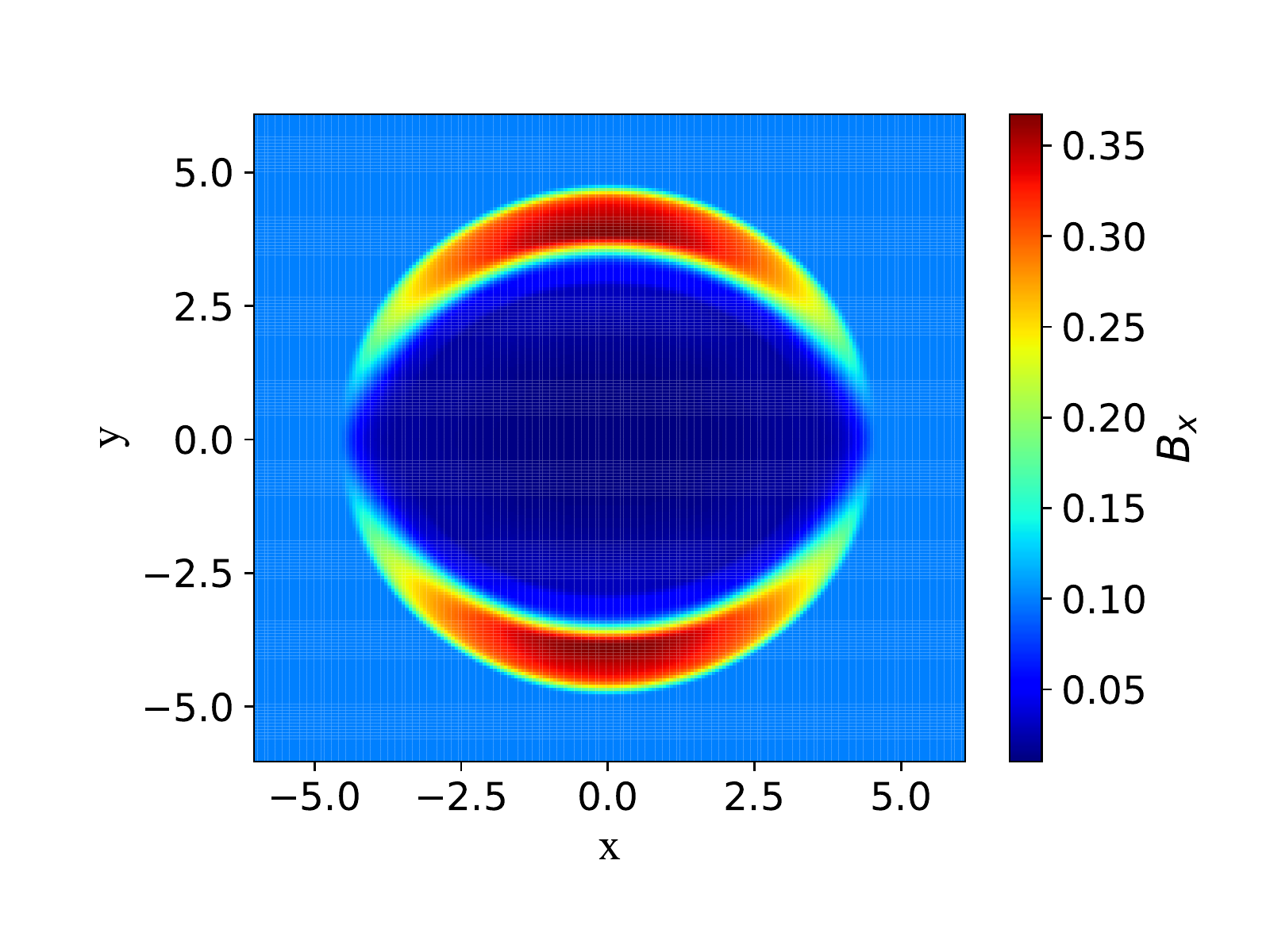}
\caption{\label{fig:CYE_Bx} The magnetic field component $B_x$ at $t = 4$ is shown in the 2D cylindrical explosion problem.}
\end{figure}

The symmetric explosions are useful standard tests for MHD codes even though there are no exact solutions because of the existence of the shock waves in all possible angles~\cite{Komissarov:2007wk}.
This problem including a strong multidimensional shock wave is useful to check the robustness of the code.
We take the same condition in Ref.~\cite{Komissarov:2007wk}.
The computational domain is $x \in [-6.0,6.0] \times y \in [-6.0,6.0]$ in the two-dimensional Cartesian coordinates. 
The number of grid points is 200 $\times$ 200 with the uniform grid.
The initial cylinder radius is taken to $r = 1$ centered at the origin, where $r = \sqrt{x^2 + y^2}$.
The pressure and density of fluid are set to $p=1$ and $\rho = 0.01$ for $r \leq 0.8$ and exponentially decrease with increasing radius for $0.8 \leq r < 1.0$.
The fluid in the exterior of the initial cylinder has $p=\rho = 0.001$ for $r \geq 1.0$.
We take the uniform magnetic field, $\bm{B} = (0.1,0.0,0.0)$, and the velocity $\bm{v} = \bm{0}$ at the initial state.
The plasma resistivity and decay constant of variables, $\psi$ and $\phi$, are set to $\eta = 1/\sigma = 0.0018$ and $\eta_d = 1/\kappa = 0.18$.
Figures~\ref{fig:CYE_P} and \ref{fig:CYE_Bx} show the two dimensional profiles at $t = 4$ of $p$ and $B_x$, respectively.
We can see that strong shock waves are formed and the initial uniform magnetic fields are deformed due to the cylindrical explosion. 
Such a deformation of the magnetic field may lead to the violation of constraints $\div{\bm{B}}=0$ and $\div {\bm{E}}=q$. If these constraints are violated, the calculation would crash due to the growth of unphysical spurious oscillation.
We can avoid these problems by adopting the GLM method. 
The results show the robustness of our code in multi-dimensional problems. We also note that our results are consistent with those of other groups (e.g., Ref.~\cite{Komissarov:2007wk}). 
We get the same results using instead $x-z$ and $y-z$ planes. 

\subsection{Rotor test}\label{RT}
The rotor test is important for a calibration of resistive as well as ideal MHD numerical multidimensional codes~\cite{DUMBSER20096991,10.1093/mnras/sts005}.
We perform the resistive rotor test both in the Cartesian and in the Milne coordinates. 
\subsubsection{Cartesian coordinates}\label{RT-Min}
\begin{figure}[ht]
\includegraphics[width=8.5cm,height=6cm]{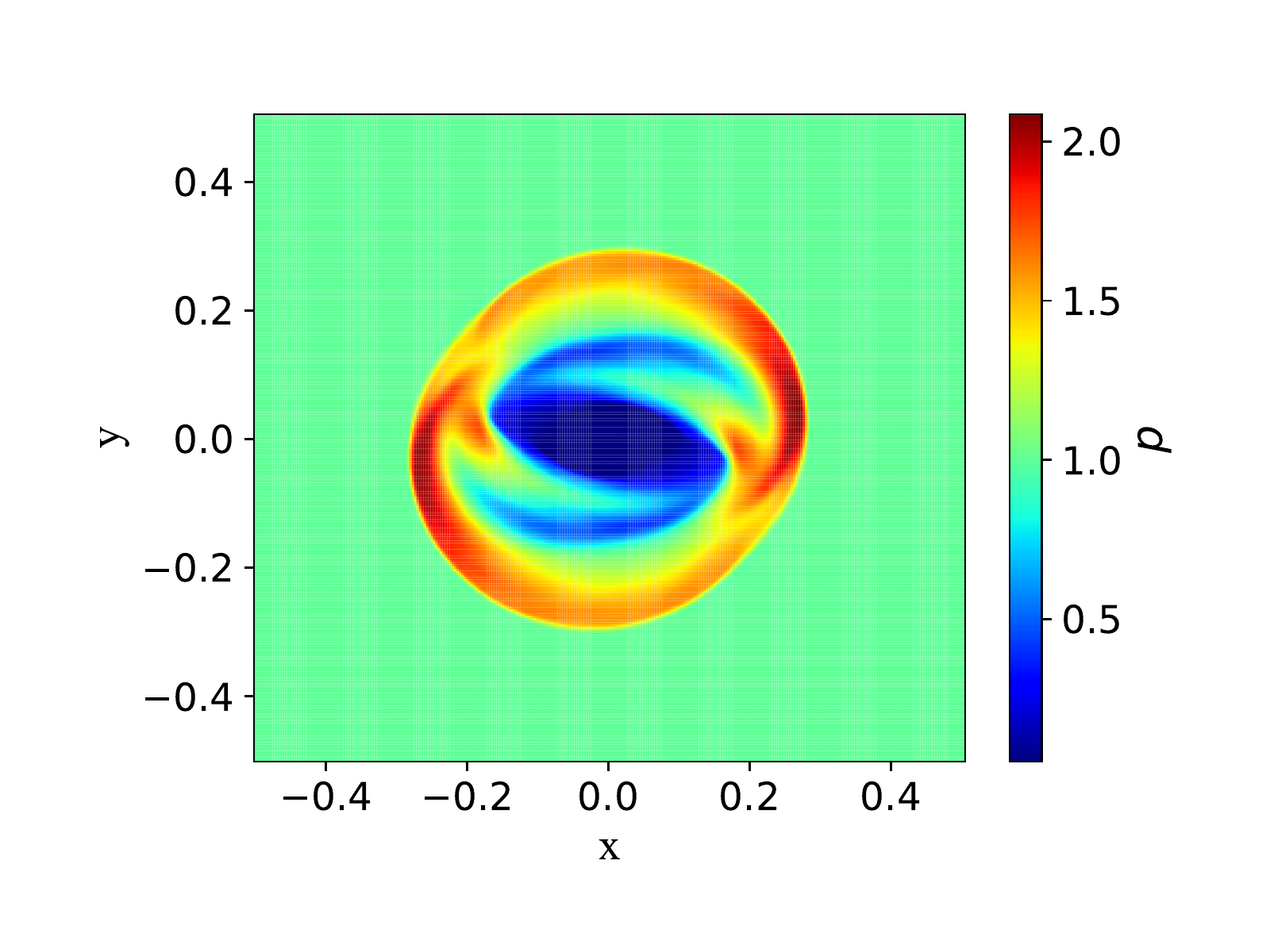}
\caption{\label{fig:RR_P} The pressure of the fluid at $t= 0.3$ is displayed in the 2D resistive rotor test in the Cartesian coordinates.}
\includegraphics[width=8.5cm,height=6cm]{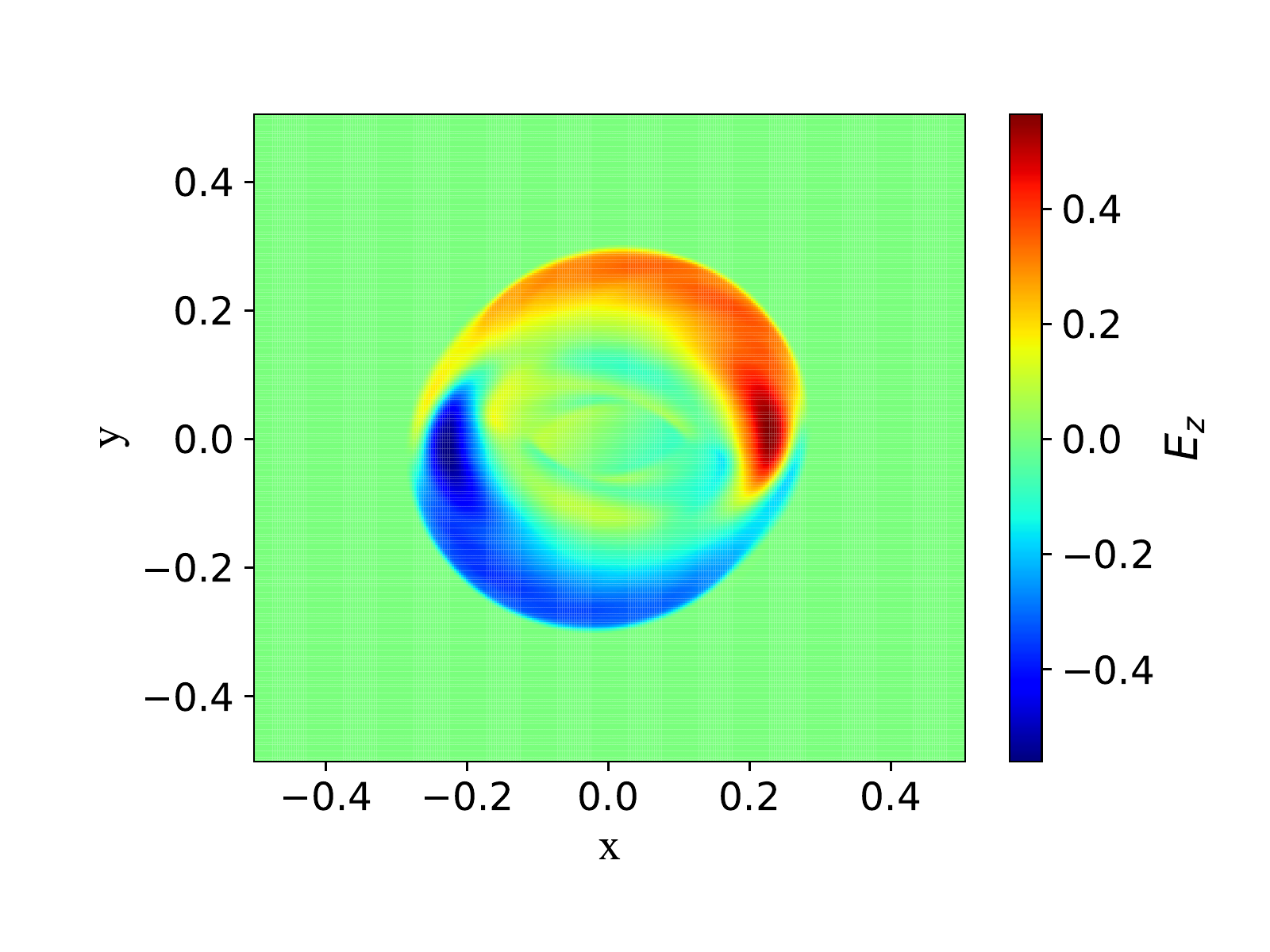}
\caption{\label{fig:RR_Ez} The electric field component $E_z$ at $t = 0.3$ is shown in the 2D resistive rotor test in the Cartesian coordinates.}
\end{figure}

The Cartesian computational domain is $x \in [-0.5,0.5] \times y \in [-0.5,0.5]$ with 300 equidistant grid points in each direction.
Inside $r \leq 0.1$ centered at the origin, the fluid rotates with constant angular velocity $\Omega = 8.5$ with uniform density $\rho = 10$ at the initial state.
Outside this region ($r \geq 0.1$), the medium is static and uniform ($\rho=1$). 
Both the pressure ($p=1$) and the magnetic field $\bm{B}=(1.0,0,0)$ are uniform in the whole region.
The initial electric field is given by the ideal condition, $-\bm{v}\times\bm{B}$.
The adiabatic index is $\Gamma = 4/3$.

Figures~\ref{fig:RR_P} and \ref{fig:RR_Ez} show snapshots of the gas pressure $p$ and the electric field component $E_z$ at $t = 0.3$ with electrical conductivity $\sigma = 10^6$.
These results agree with that of other simulation code~\cite{DUMBSER20096991,10.1093/mnras/sts005,10.1093/mnras/sty419}. 
Again, the GLM method does work in our code, so that the growth of unphysical oscillation suppresses.
The same results are reproduced by taking $x-z$ and $y-z$ planes.

\subsubsection{Milne coordinates}\label{RT-Milne}

\begin{figure*}[t]
  \begin{minipage}[l]{0.45\linewidth}
    \includegraphics[width=8cm,height=6cm]{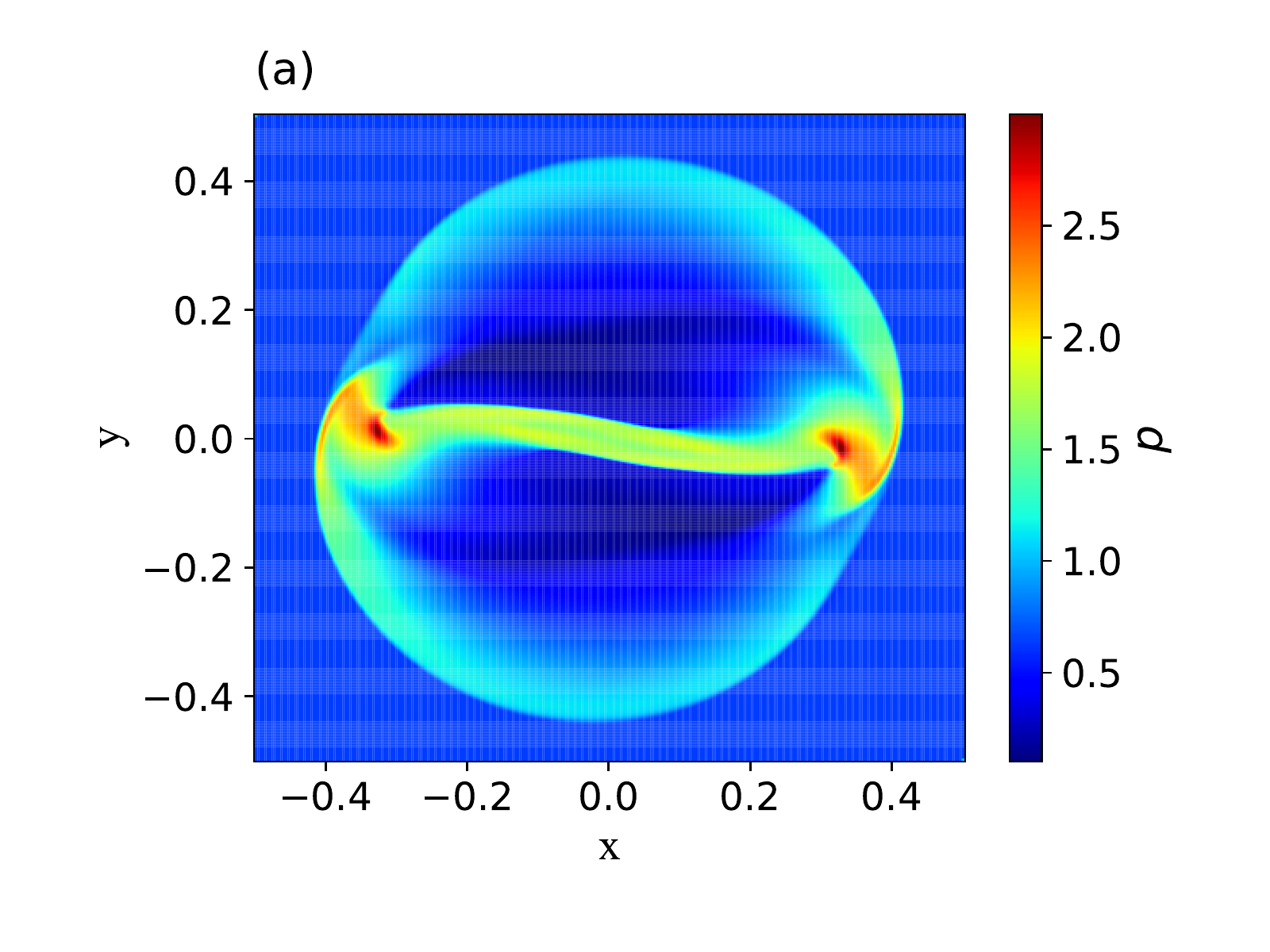}
  \end{minipage}
  \begin{minipage}[r]{0.45\linewidth}
    \includegraphics[width=8cm,height=6cm]{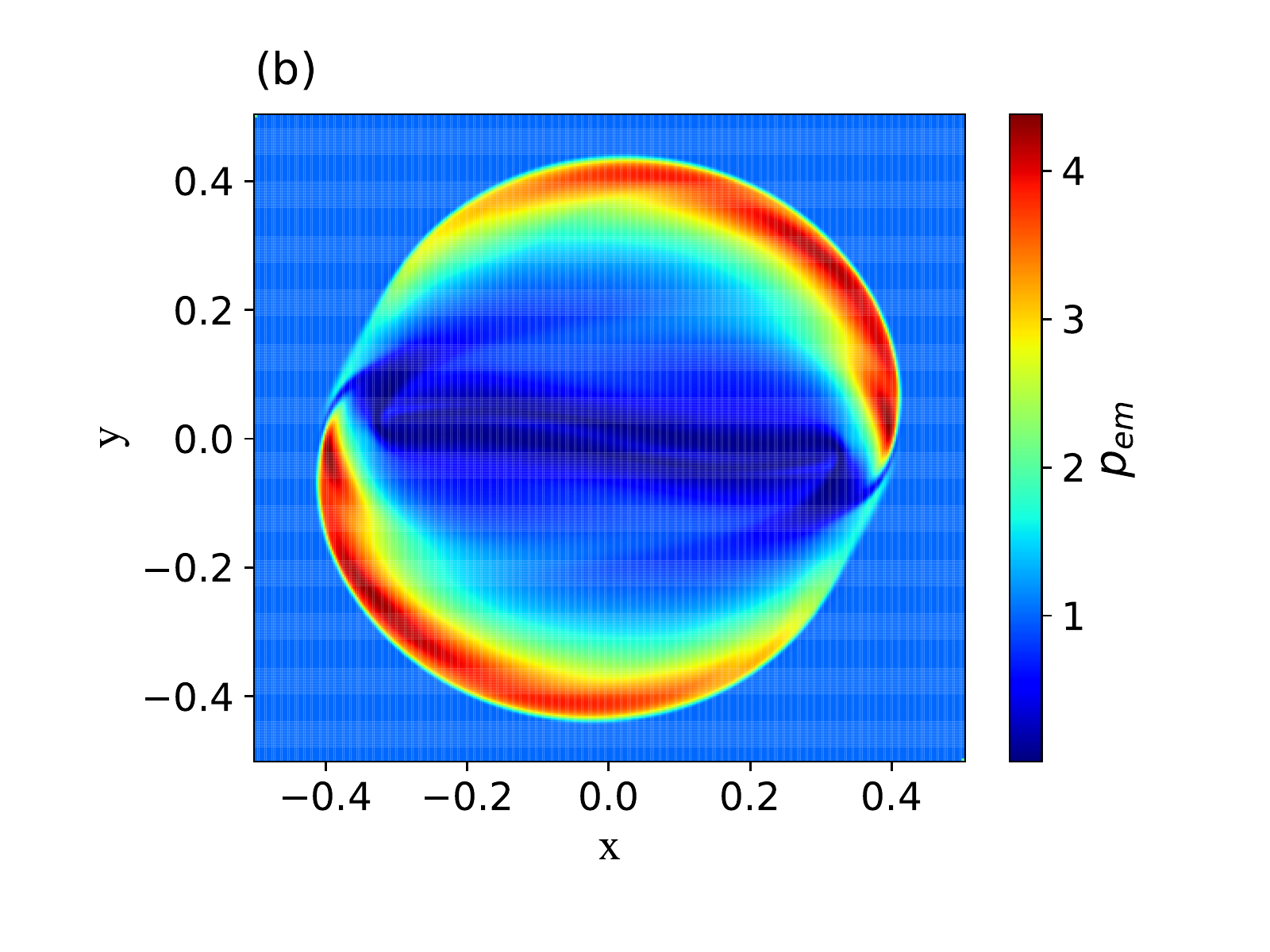}
     \end{minipage}
    \caption{(color online) (a) The pressure of fluid and (b) the energy density of the electromagnetic fields at $t = 1.4$ are shown in the 2D resistive rotor test in the Milne coordinates in the case of $\sigma = 10^3$.}
    \label{fig:RRM_sig10e3}
\end{figure*}

\begin{figure*}[t]
  \begin{minipage}[l]{0.45\linewidth}
    \includegraphics[width=8cm,height=6cm]{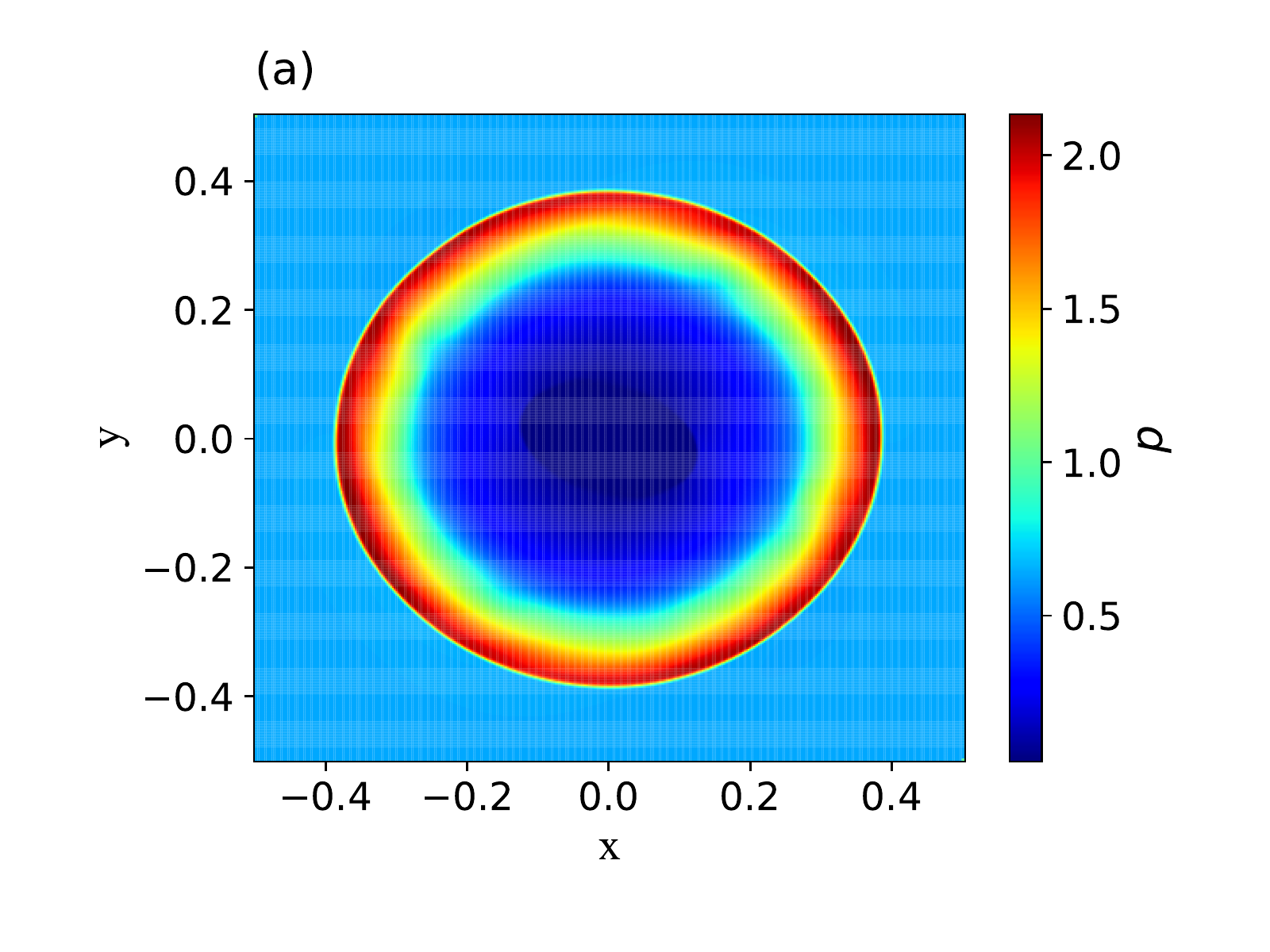}
  \end{minipage}
  \begin{minipage}[r]{0.45\linewidth}
    \includegraphics[width=8cm,height=6cm]{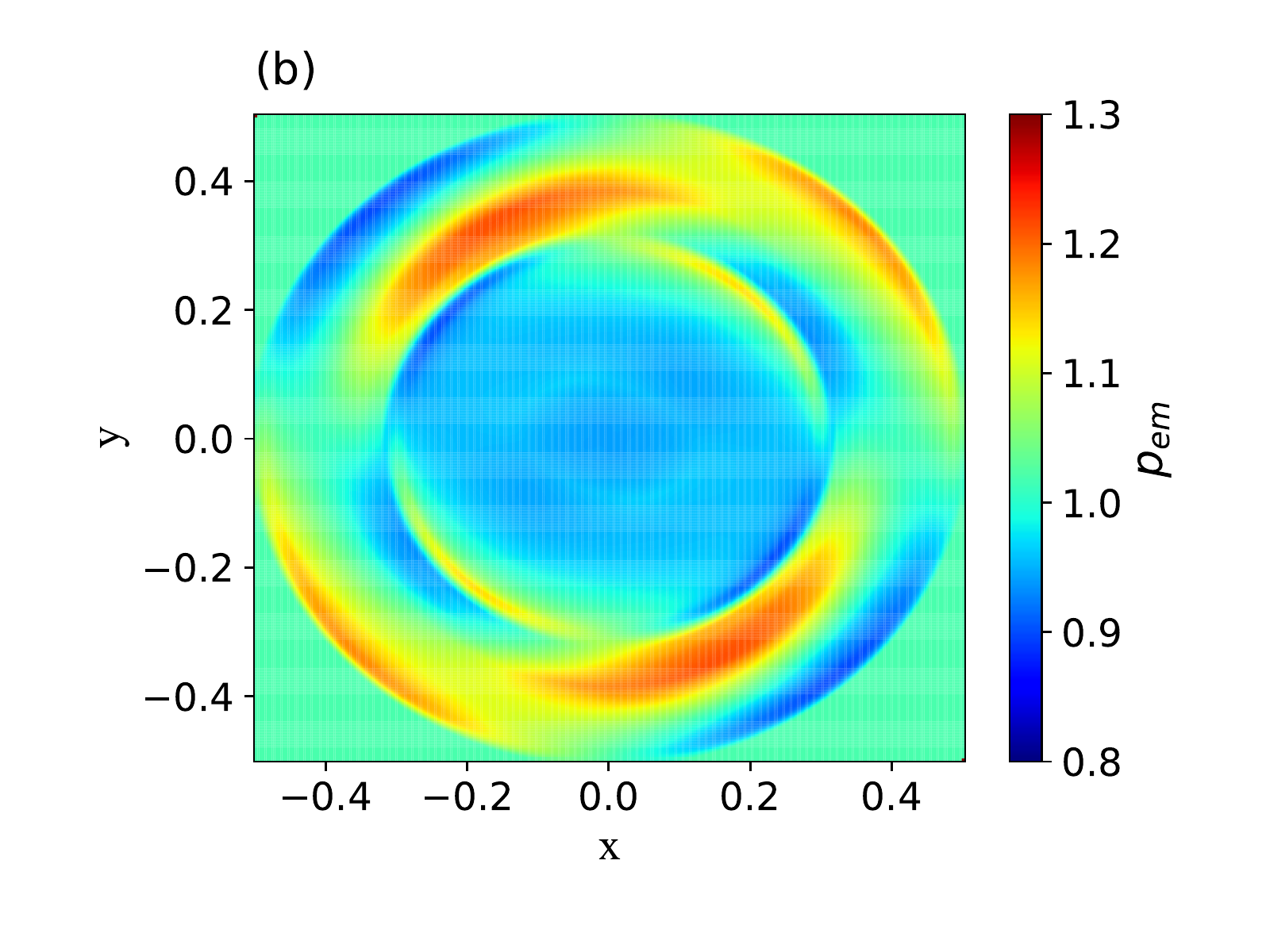}
     \end{minipage}
    \caption{(color online) (a) The pressure of fluid and (b) the energy density of the electromagnetic fields at $t = 1.4$ are shown in the 2D resistive rotor test in the Milne coordinates in the case of $\sigma = 10$.}
    \label{fig:RRM_sig10e1}
\end{figure*}
The rotor test problem for the relativistic ideal MHD in the Milne coordinates is proposed in Ref.~\cite{Inghirami:2016iru}.
The computational domain in the Milne coordinates is $x \in [-0.5,0.5] \times y \in [-0.5,0.5]$ with 400 equidistant grid points in each direction.
We adopt the free boundary condition in both directions.
In this test, the EoS is assumed to be the ultra-relativistic ideal gas EoS, $p = e/3$, and the fluid density is set to $\rho = 0$ at all grid points.
We set the initial coordinate time to $t = 1.0$.
Instead of imposing high density around the origin, we set a high pressure region $p = 5$ inside $r=0.1$.
The initial transverse velocity $(v_x,v_y)$ is given by angular speed $\Omega=9.7$.
Outside this region, the pressure is $p=1$.
The initial longitudinal velocity is set to $v_\eta = 0$, which is assumed to be longitudinal Bjorken expansion amount to $v^z = z/t$.
The initial magnetic field is set to $\bm{B} = (2.0,0,0)$ and the electric field is taken to be $-\bm{v}\times\bm{B}$.

Figures~\ref{fig:RRM_sig10e3} (a) and (b) show the results of $p$ and $p_{em}$ of our simulation code  at $t = 1.4$ with $\sigma = 10^3$, respectively.
In the Milne coordinates, we observe the decay of the pressure of the fluid and magnetic pressure, which occurs in the whole computational domain because of the longitudinal expansion of the system.
An asymmetric shaped compression wave by the magnetic field is observed in our results because of the higher initial pressure inside the radius $r < 0.1$ and the rotation of the cylinder. 
Our result with $\sigma = 10^3$ has a similar configuration of the result in ECHO-QGP simulation~\cite{Inghirami:2016iru}.
In the high conductive case, our results capture the features of relativistic ideal MHD.
Figures~\ref{fig:RRM_sig10e1} (a) and (b) represent the results of $p$ and $p_{em}$ of our simulation code  at $t = 1.4$ with $\sigma = 10$, respectively.
We can see that the anisotropy becomes weaker for a lower conductivity.
It indicates that the fluid is weakly coupled with electromagnetic fields.
The fluid vorticity does not affect the dynamics of electromagnetic fields.
These features can be captured in our code.

\subsection{Magnetized Bjorken flow}\label{Bjorken}

This problem is an extension of the one-dimensional boost invariant flow proposed by Bjorken~\cite{PhysRevD.27.140} to the ideal MHD~\cite{ROY201545}.
We consider the relativistic boost invariant flow in the $z$-direction.
In magnetized Bjorken flow, we consider that the fluid follows the ultra-relativistic ideal gas EoS, $p = e/3$.
The pressure of fluid is assumed to be uniform in space. 
The fluid velocity is given by $v_z=z/t$, which is obtained by assuming a longitudinal boost invariant expansion. 
The corresponding four-velocity in the Cartesian coordinates is given by $u^\mu = (\cosh{\eta_s},0,0,\sinh{\eta_s})$.
In the Milne coordinates, the four-velocity becomes simply $u^\mu = (1,0,0,0)$.
The comoving derivative and the expansion rate become $D = \partial_\tau$ and $\Theta = 1/\tau$, respectively. 
The magnetic field is given by $b^\mu = (0,b^x,b^y,0)$.
In this assumption, from the energy conservation equation Eq.~(\ref{emc}), one derives~\cite{ROY201545}, 
\begin{equation}
    \partial_{\tau}\left(e+\frac{b^2}{2}\right) + \frac{e+p+b^2}{\tau} = 0,
\end{equation}
where $b^2 = b^\mu b_\mu$.
From the energy equation Eq.~(\ref{energy conservation equation}), we obtain, 
\begin{equation}
    \partial_\tau e + \frac{e+p}{\tau} = 0,
\end{equation}
for the ideal MHD.
The evolution equation of the magnetic field is obtained as,
\begin{equation}
    \partial_\tau b + b/\tau = 0.
\end{equation}

The analytic solutions of these equations for the ultra-relativistic ideal gas EoS, $p = e/3$, are given by,
\begin{gather}
    e(\tau) = e_0\left(\frac{\tau_0}{\tau}\right)^{4/3},\\
    b(\tau) = b_0\frac{\tau_0}{\tau},
\end{gather}
where $\tau_0 = 0.5$ fm is initial time, $e_0 = 10~\mathrm{GeV/fm^3}$ is initial energy density, and $b_0 = 1.0 ~\mathrm{GeV^{1/2}/fm^{3/2}}$ is initial magnetic field. 

\begin{figure}[ht]
\includegraphics[width=8.5cm,height=6cm]{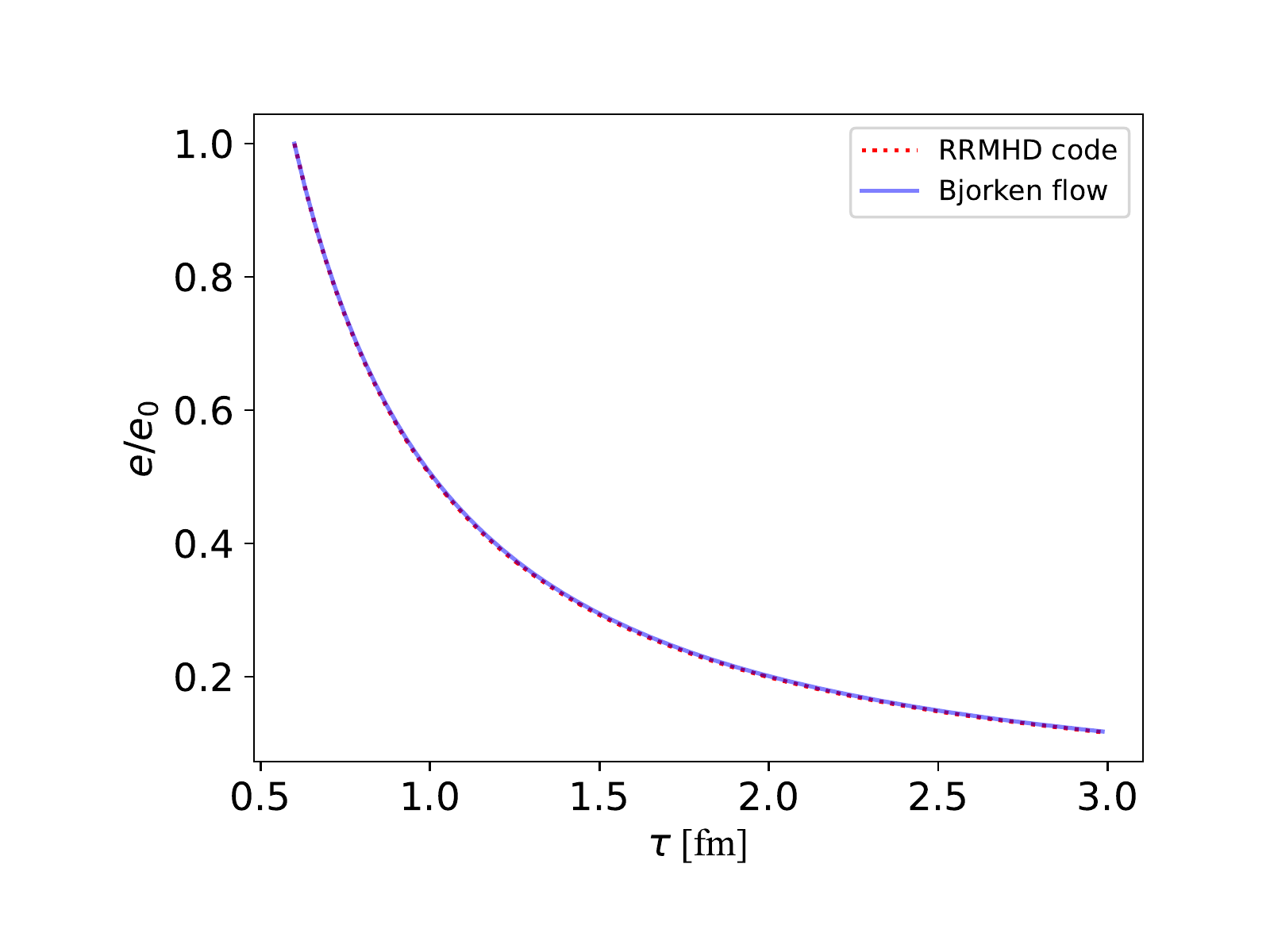}
\caption{\label{fig:Bjorken_e} (color online) The energy density of the fluid is displayed as a function of $\tau$ in the magnetized Bjorken flow.}
\end{figure}

\begin{figure}[ht]
\includegraphics[width=8.5cm,height=6cm]{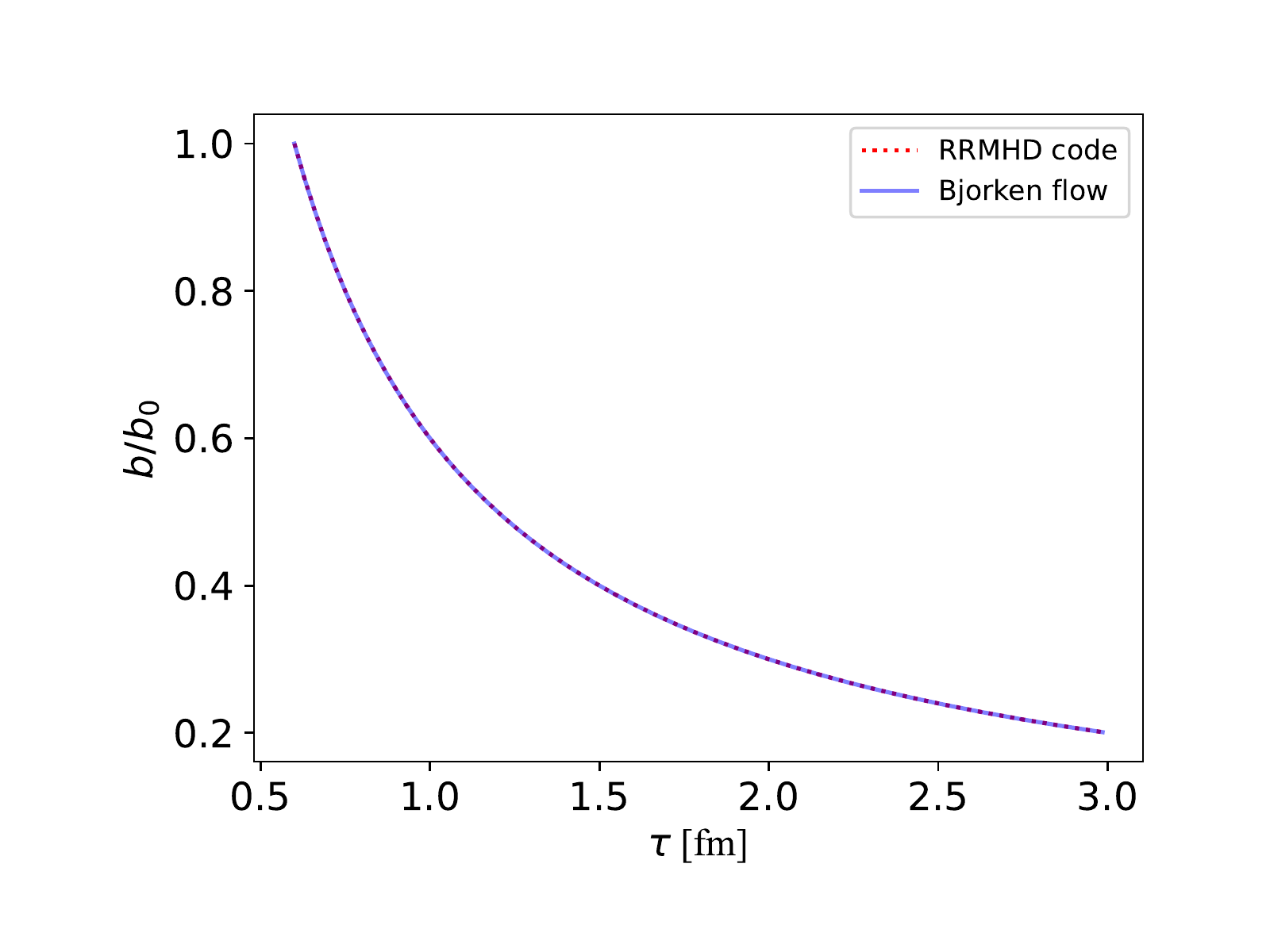}
\caption{\label{fig:Bjorken_b} (color online) The magnetic field strength as a function of $\tau$ is displayed in the magnetized Bjorken flow.}
\end{figure}

Figures~\ref{fig:Bjorken_e} and \ref{fig:Bjorken_b} show the time evolution of energy density and magnetic field strength with $\sigma = 10^6$, respectively. The solid curves show analytic solutions, while the dashed curves denote the numerical results.
These results are in good agreement with the analytic solutions.

\subsection{Accelerating longitudinal expansion}\label{ALE}

\begin{figure*}[t]
  \begin{minipage}[l]{0.45\linewidth}
    \includegraphics[width=8cm,height=6cm]{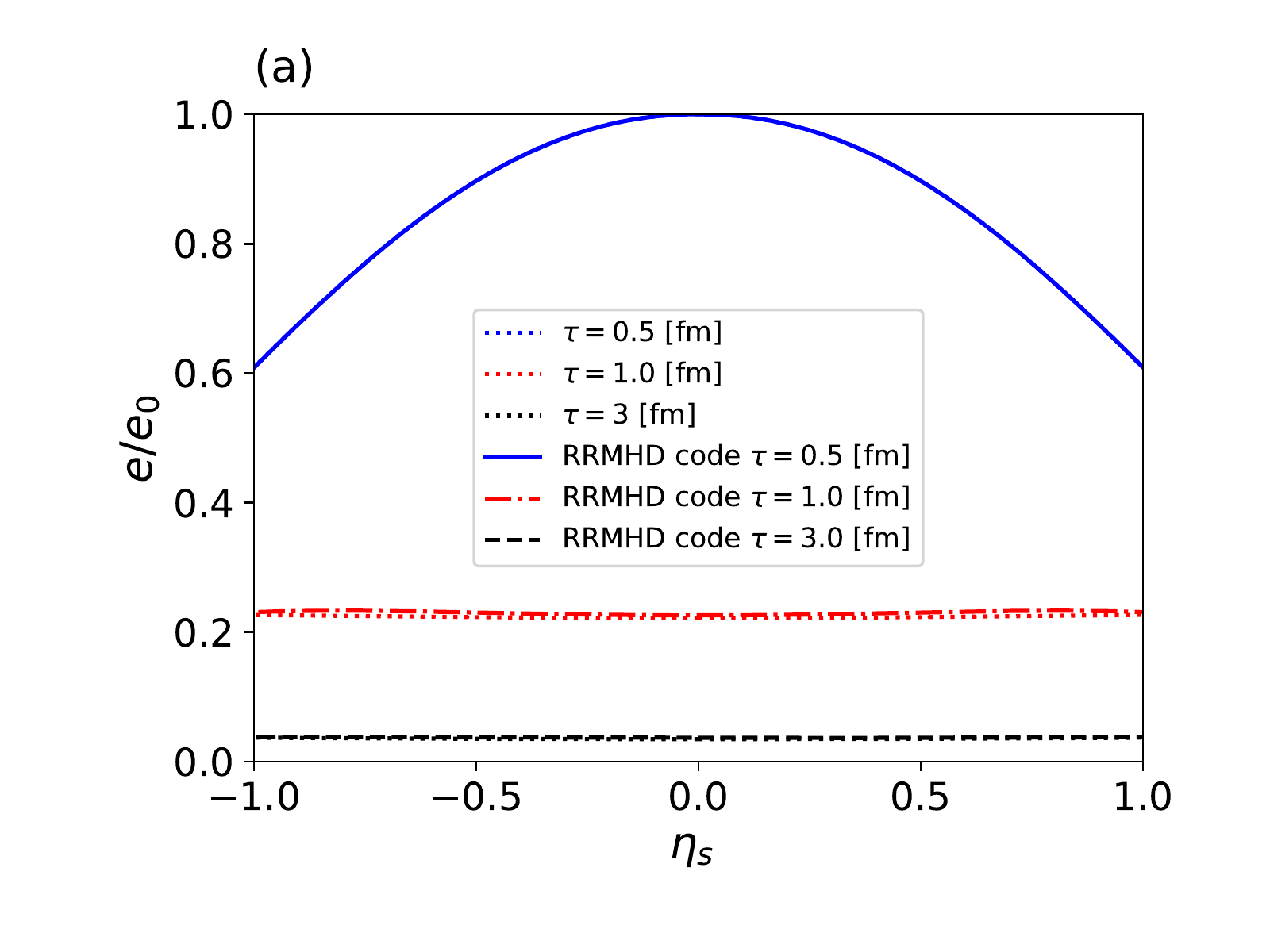}
  \end{minipage}
  \begin{minipage}[r]{0.45\linewidth}
    \includegraphics[width=8cm,height=6cm]{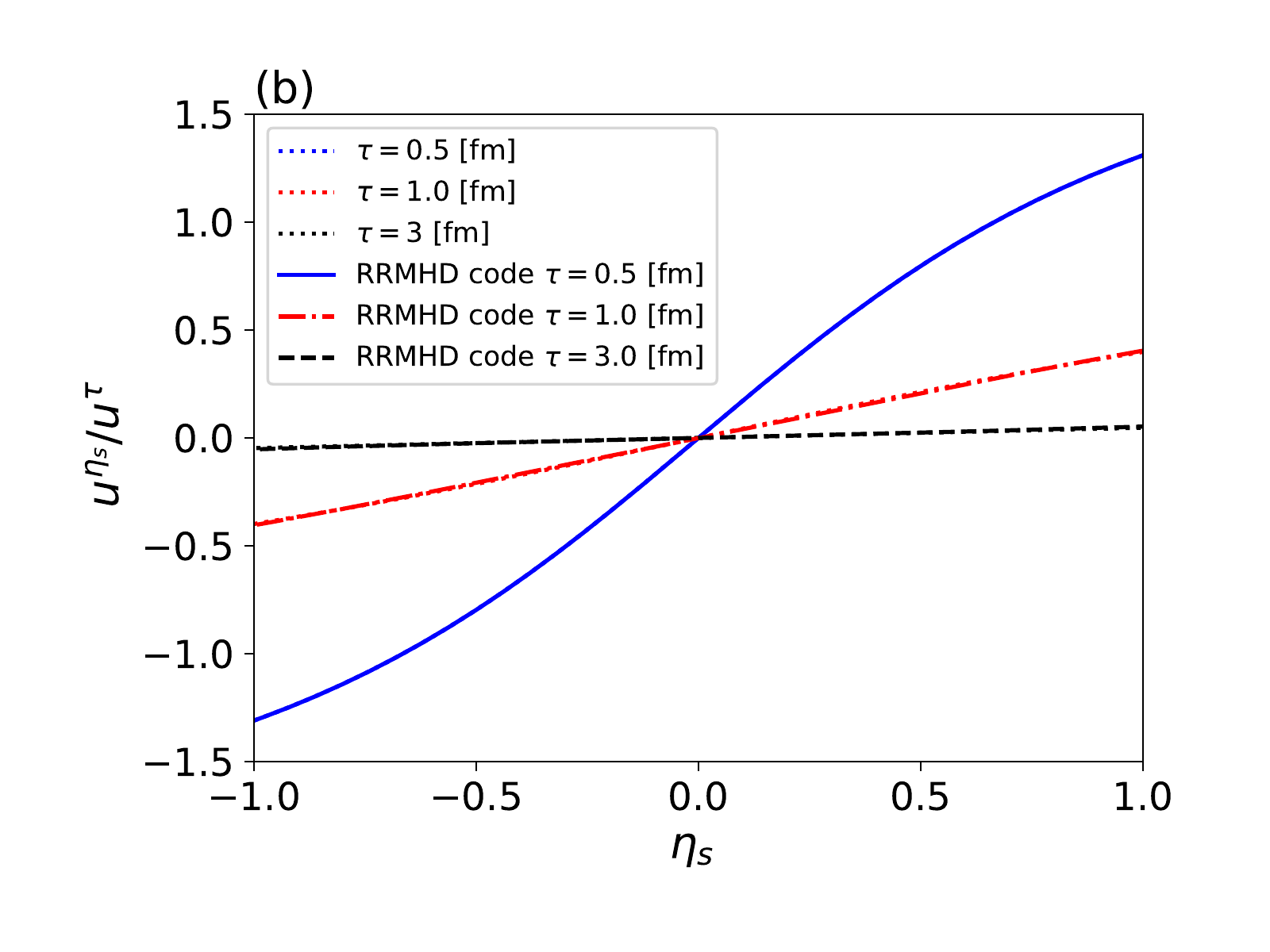}
     \end{minipage}
    \caption{(color online) We display (a) the ratio of the energy density of the fluid to the initial energy density and (b) the fluid velocity component $v^\eta$, respectively.
    The blue solid, red long dashed-dotted and black dashed lines show the numerical results at $t = 0.5, 1.0$, and $3$ fm, respectively.
    The blue, red, and black dotted lines show the semi-analytic solutions of the Eqs.~(\ref{ALE_mm}) and (\ref{ALE_mm2}) at $t = 0.5, 1.0$ and $3.0$ fm, respectively.}
    \label{fig:ALE_fluid}
\end{figure*}

This test problem is presented in Ref.~\cite{PhysRevD.102.014017}.
This is a resistive extension of the magnetized Bjorken flow~\cite{ROY201545}.
The derivation of the semi-analytic solutions is shown in Appendix~\ref{Appendix:ALE} in detail, but we here show the results.
In this problem, we do not suppose the boost invariant flow.
We assume that the fluid velocity is parallel to the longitudinal ($z$-) direction, and the fluid is uniform on the transverse ($x$-$y$) plane.
Let us parametrize the four-velocity in the Milne coordinates as follows:
\begin{eqnarray}
  u^\mu &=& \left(\cosh (Y-\eta_s),0,0,\frac{1}{\tau}\sinh(Y-\eta_s)\right)\label{fluid velocity in ALE_M}\\
        &=&\bar{\gamma}\left(1,0,0,\frac{1}{\tau}\bar{v}\right),\nonumber
\end{eqnarray}
where $Y$ is the rapidity with $\bar{\gamma} = \cosh(Y-\eta_s)$ and $\bar{v} = \tanh(Y-\eta_s)$.

The electric and magnetic four vectors are considered on the transverse plane and orthogonal to each other,
\begin{gather}
    e^\mu = (0,e^x,0,0),\nonumber\\
    b^\mu = (0,0,b^y,0).\label{electromagnetic field in ALE_M}
\end{gather}
In Ref.~\cite{PhysRevD.102.014017}, electromagnetic fields are taken to be the following forms,
\begin{gather}
    e_x(\tau,\eta_s) = - \frac{c(\eta_s)}{\tau}\sinh(Y-\eta_s),\\
    b_y(\tau,\eta_s) =  \frac{c(\eta_s)}{\tau}\cosh(Y-\eta_s).
\end{gather}
The solutions for the rapidity and electromagnetic fields are given by,
\begin{gather}
    Y = \eta_s + \sinh^{-1}\left(\frac{1}{\sigma\tau}\frac{\partial_{\eta_s}c(\eta_s)}{c(\eta_s)}\right),\label{ALE_rapidity}\\
    u^{\tau} = \sqrt{1+\left(\frac{1}{\sigma\tau}\frac{\partial_{\eta_s}c(\eta_s)}{c(\eta_s)}\right)^2},\\
    u^{\eta_s} = \frac{1}{\sigma\tau^2}\frac{\partial_{\eta_s}c(\eta_s)}{c(\eta_s)},\label{ALE_veta_M}\\
    e_x(\tau,\eta_s) = \frac{1}{\sigma\tau^2}\frac{\partial c(\eta_s)}{\partial\eta_s},\\
    b_y(\tau,\eta_s) = \frac{c(\eta_s)}{\tau}\times\sqrt{1+\left(\frac{1}{\sigma\tau}\frac{\partial_{\eta_s}c(\eta_s)}{c(\eta_s)}\right)^2}\label{ALE_magneticfield}.
\end{gather}
Here $c(\eta_s)$ is an arbitrary function. In this paper, we suppose,
\begin{equation}
    c(\eta_s) = c_0\cosh(\alpha\eta_s)\label{ALE_c},
\end{equation}
which is the same as that in Ref.~\cite{PhysRevD.102.014017}.
Here $\alpha$ is an arbitrary constant. 
The arbitrary constant $c_0$ is taken to be $0.0018\tau_0$, which determines the initial laboratory frame magnetic field strength in the Minkowski coordinates $B_L^y(\tau_0, 0) = 0.0018~\mathrm{GeV^2}/e$.

The energy density is determined by solving the following equations,
\begin{gather}
    \partial_\tau e(\tau,\eta_s) + \frac{1+\kappa}{\tau}A(\tau,\eta_s)e(\tau,\eta_s) = B(\tau,\eta_s),\label{ce_ALE1_M}\\
    \partial_{\eta_s}e(\tau,\eta_s)+H(\tau,\eta_s)e(\tau,\eta_s) = G(\tau,\eta_s),\label{ce_ALE2_M}
\end{gather}
where,
\begin{gather}
    A(\tau,\eta_s) = \left(\frac{\partial_{\eta_s}Y(\bar{v}-\kappa)-(\kappa-1)\tau\bar{v}\partial_\tau Y}{\kappa(\bar{v}^2-1)}\right)\label{ALE_A}\\
    B(\tau,\eta_s) = \frac{\sigma(e_x b_y\bar{v}-\kappa e_x^2)}{\kappa\bar{\gamma}(\bar{v}^2-1)},\\
    H(\tau,\eta_s) = \frac{1}{\kappa}((1+\kappa)(\tau\partial_\tau Y+\bar{v}\partial_{\eta_s}Y))-(1+\kappa)\bar{v}A(\tau,\eta_s),\\
    G(\tau,\eta_s) = \frac{(\sigma\tau)e_x b_y}{\bar{\gamma}\kappa}-\tau\bar{v}B(\tau,\eta_s)\label{ALE_G}.
\end{gather}
Here, we take the ideal gas EoS, $p = \kappa e$ and $\kappa = 1/3$.
We numerically solve these differential equations on a grid of points on the ($\tau,\eta_s$) plane.
First, we give the arbitrary function $c(\eta_s)$, which is assumed to have a form given in Eq.~(\ref{ALE_c}). Then the rapidity, four-velocity, and electromagnetic fields are obtained from Eqs.~(\ref{ALE_rapidity})-(\ref{ALE_magneticfield}).
The function $A(\tau,\eta_s)$, $B(\tau,\eta_s)$, $H(\tau,\eta_s)$, and $G(\tau,\eta_s)$ are determined from Eqs.~(\ref{ALE_A})-(\ref{ALE_G}). 
We solve Eq.~(\ref{ce_ALE1_M}) as an ordinary differential equation (ODE) to find out the $\tau$-dependence of the function $e$, keeping constant as the variable $\eta_s$.
Then, we solve Eq.~(\ref{ce_ALE2_M}) with the solution of Eq.~(\ref{ce_ALE1_M}) in each $\tau$ as an initial condition of the ODE.
The energy density is obtained as numerical solutions of two ODEs with the initial value of the $e(\tau_0,0) = e_0$.

Here, we perform the RRMHD simulation using the numerical solution of the above ODEs as the initial condition.
Since the resistive effect is prominent with small $\alpha$~\cite{PhysRevD.102.014017}, we take the initial energy density $e_0=1.0~\mathrm{GeV/fm^3}$, $\alpha = 0.1$, and the electrical conductivity $\sigma = 0.023~\mathrm{fm^{-1}}$.
The number of grid points in the computational domain $\eta_s \in [-3.0,3.0]$ is $N = 200$.
We adopt free boundary conditions at $\eta_s = \pm 3.0$.
The waves are, however, sometimes reflected at the boundaries and they affect the numerical results.
To avoid this, we take the boundaries far away from the central region $\eta_s \in [-1.0,1.0]$ and stop the calculation before the reflected waves reach the central region. 

\begin{figure*}[t]
  \begin{minipage}[l]{0.45\linewidth}
    \includegraphics[width=8cm,height=6cm]{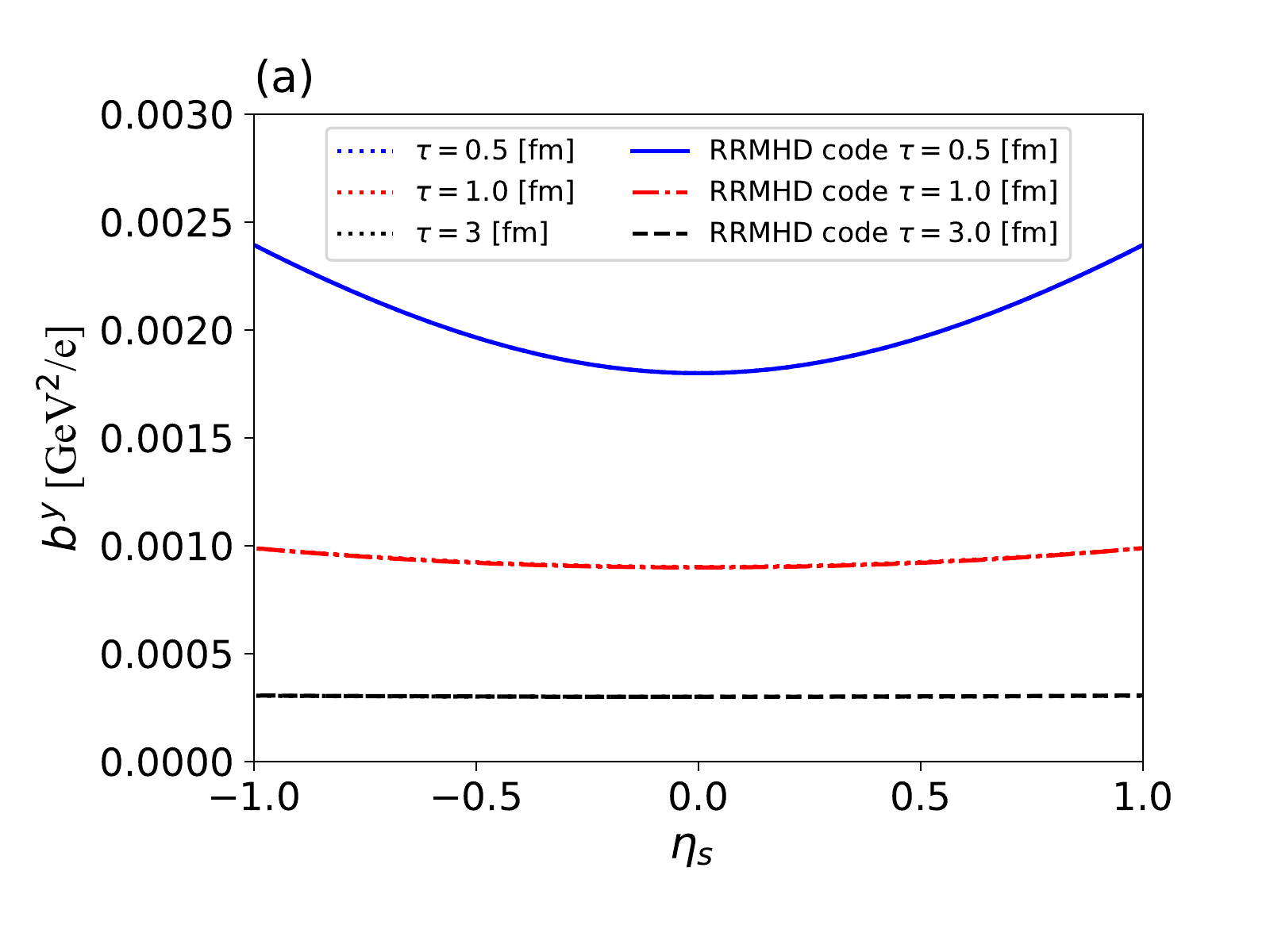}
  \end{minipage}
  \begin{minipage}[r]{0.45\linewidth}
    \includegraphics[width=8cm,height=6cm]{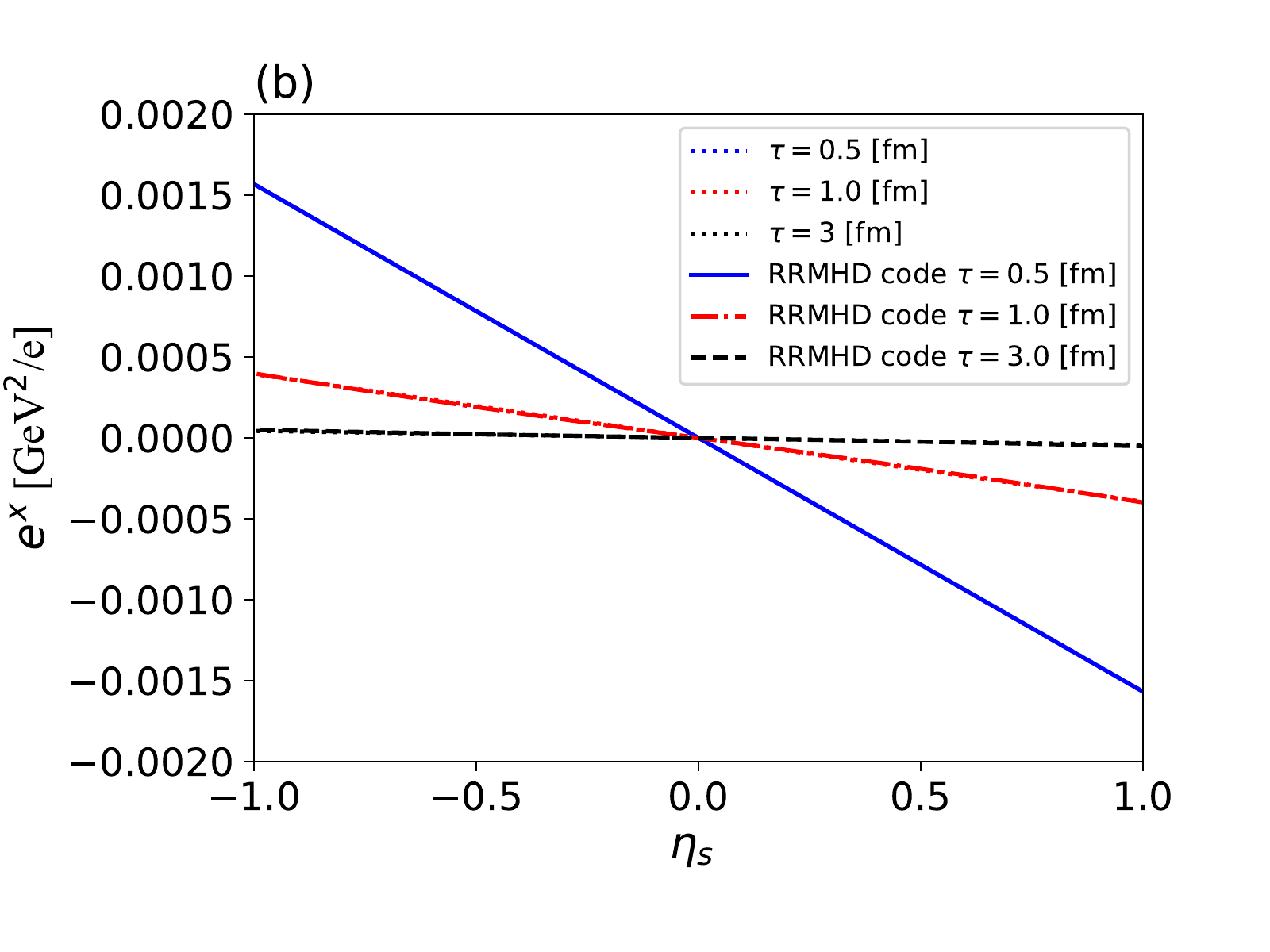}
     \end{minipage}
    \caption{(color online) We display (a) the magnetic field component $b^y$ and (b) the electric field component measured in the comoving frame $e^x$, respectively.
    The blue solid, red long dashed-dotted and black dashed lines show the numerical results at $t = 0.5, 1.0$, and $3$ fm, respectively.
    The blue, red, and black dotted lines show the semi-analytic solutions of the Eqs.~(\ref{ALE_mm}) and (\ref{ALE_mm2}) at $t = 0.5, 1.0$ and $3.0$ fm, respectively.
    }
    \label{fig:ALE_EM}
\end{figure*}
Figures~\ref{fig:ALE_fluid} and \ref{fig:ALE_EM} show the results of our simulation code in comparison with the semi-analytic solution.
Our results are in good agreement with the semi-analytic solutions.
In Fig.~\ref{fig:ALE_fluid} (a), the energy density decays and expands to the longitudinal direction by the resistive effects.
In Fig.~\ref{fig:ALE_fluid} (b), the fluid velocity in the Milne coordinates has a finite value in the forward and backward rapidity regions and it decays with time.
This feature comes from the resistive effect described in Eq.~(\ref{ALE_veta_M}).
Our results reproduce this behavior of the fluid.
In Fig.~\ref{fig:ALE_EM} (a), the magnetic field component in the comoving frame $b^y$ as a function of $\eta_s$ is represented.
The magnetic field decays with time by the longitudinal expansion similar to the magnetic Bjorken flow in Subsection.~\ref{Bjorken}.
However, the magnetic field has a non-uniform profile in the $\eta_s$ direction by the acceleration induced from the resistive effect, which is different from the magnetic Bjorken flow.   
Figure~\ref{fig:ALE_EM} (b) shows the electric field component measured in the coming frame $e^x$ as a function of $\eta_s$.
The electric field has a positive value in the backward rapidity region and it decreases with rapidity.
The electric field changes its sign at $\eta_s = 0$.
This feature describes that the electric field is produced by the two colliding nuclei in high-energy heavy-ion collisions.
Our results capture these features and its diffusion which is consistent with the semi-analytic solutions.





\section{Summary and Discussion}\label{summary}
We constructed a new RRMHD simulation code in the Milne coordinates which is suitable for study on high-energy heavy-ion collisions.
We split the system of RRMHD equations into two parts, a non-stiff and a stiff part.
The primitive variables were interpolated from the cell center to the cell surface by using the second-order accurate scheme~\cite{VANLEER1977276}.
For the non-stiff part, we evaluated the numerical flux using the HLL approximated Riemann solver and explicitly integrated the equations in time by the second-order of Runge-Kutta algorithm~\cite{SHU1988439}.
For the stiff part appeared in Ampere's law, we executed time integration using semi-analytic solutions to avoid unexpected small time steps.
Though Maxwell equations ensure that the divergence-free constraints are satisfied at all times, in numerical simulation, the integration of Maxwell equations in a not well-designed scheme does not preserve these conditions because of the numerical error.
In order to avoid this problem, we employed the generalized Lagrange multiplier method to guarantee these conditions~\cite{MUNZ2000484,Komissarov:2007wk,Porth:2016rfi}.

We checked the correctness of our algorithm from the comparison between numerical calculations and analytical solutions or the other RMHD simulations such as Brio-Wu type shock tubes, propagation of the large amplitude circularly polarized Alf\'{v}en waves, self-similar current sheet, cylindrical explosion, resistive rotor, and Bjorken flow.
In these test problems, our numerical solutions were in good agreement with analytic solutions or results of the other RRMHD simulations.
Furthermore, we investigated the accelerating longitudinal expansion of relativistic resistive magneto-hydrodynamics in high-energy heavy-ion collisions in comparison with semi-analytic solutions~\cite{PhysRevD.102.014017}.
Our numerical code reproduced these solutions.
We conclude that our numerical simulations capture the characteristic features of dynamics in high-energy heavy-ion collisions.

We shall employ our RRMHD code to investigate experimental data at RHIC and the LHC and understand the detailed QGP bulk property such as electrical conductivity.
As the first application, we have demonstrated calculation of the directed flow in RRMHD expansion in Au-Au and Cu-Au collision systems at RHIC~\cite{Nakamura:2022idq}.
We found that electrical conductivity prevents the growth of the directed flow in asymmetric systems. 
The Ohmic conduction current may affect the charge-dependent flow of hadrons.
Recently, the charge-dependent directed flow of the heavy mesons and dileptons has been proposed as a signature of the electromagnetic fields~\cite{SUN2021136271,PhysRevC.105.054907}. 
It can extract the initial electromagnetic fields and the QGP bulk properties from experimental results.

The highly intense electromagnetic fields induce novel quantum phenomena such as the chiral magnetic effect (CME) ~\cite{PhysRevD.78.074033,KHARZEEV2008227} and the chiral magnetic wave (CMW)~\cite{PhysRevD.83.085007}.
The CME and CMW signals in experimental data have been explored in iso-bar collisions, Zr-Zr and Ru-Ru collisions at RHIC~\cite{ZHAO2019200,PhysRevC.105.014901}.
However, there are no striking evidences of these phenomena in the iso-bar experiments. 
For analytical studies, the initial condition of electromagnetic fields with CME in high-energy heavy-ion collisions has been studied~\cite{PhysRevC.91.064902,PhysRevC.105.054909}, but the evolution of these fields has not been well understood. The comprehensive studies using RRMHD would shed light on these problems. 
This issue remains an important future work.



\section*{Acknowledgement}
The work of K.N. was supported in part by JSPS Grant-in-Aid for JSPS Research Fellow No.~JP21J13651.
This work was also supported by JSPS KAKENHI Grant Numbers, JP19H01928, JP20K11851 (T.M.), 
JP20H00156, JP20H11581, JP17K05438 (C.N.), JP20H00156, JP20H01941, JP20K11851, and JP21H04488 (H.R.T.).

\appendix

\section{Solution of accelerating longitudinal expansion}\label{Appendix:ALE}
We show details of the test problem in Subsection~\ref{ALE}, which is proposed in Ref.~\cite{PhysRevD.102.014017}.
This is a resistive extension of the magnetized Bjorken flow~\cite{ROY201545}.
In this problem, we do not assume the boost invariant flow.
However, we suppose that the fluid velocity keeps parallel to the longitudinal direction while the transverse flow is neglected.
All the fluid quantities are uniform in the transverse plane.
Let us parametrize the four-velocity in (1+1)D as follows:
\begin{equation}
    u^\mu = \gamma(1,0,0,v_z) = (\cosh Y,0,0,\sinh Y),
\end{equation}
where $Y$ is the rapidity and $v_z = \tanh Y$.
In the Milne coordinates, $u^\mu$ is rewritten by,
\begin{eqnarray}
  u^\mu &=& \left(\cosh (Y-\eta_s),0,0,\frac{1}{\tau}\sinh(Y-\eta_s)\right)\label{fluid velocity in ALE}\\
        &=&\bar{\gamma}\left(1,0,0,\frac{1}{\tau}\bar{v}\right),\nonumber
\end{eqnarray}
where $\bar{\gamma} = \cosh(Y-\eta_s)$ and $\bar{v} = \tanh(Y-\eta_s)$.
Under this parameterization, the comoving derivative and the expansion rate are given by,
\begin{gather}
    D = \bar{\gamma}\left(\partial_\tau + \frac{1}{\tau}\bar{v}\partial_{\eta_s}\right),\\
    \Theta = \bar{\gamma}\left(\bar{v}\partial_\tau Y+\frac{1}{\tau}\partial_{\eta_s} Y\right).
\end{gather}
Then, Eq.~(\ref{energy conservation equation}) becomes,
\begin{equation}
    (\tau\partial_\tau + \bar{v}\partial_{\eta_s})e + (e+p)(\tau\bar{v}\partial_\tau Y + \partial_{\eta_s} Y)=\bar{\gamma}^{-1}\tau e^\lambda J_{\lambda},\label{energy}
\end{equation}
and Eq.~(\ref{momentum conservation equation}) in the case of $\alpha = \eta_s$ gives,
\begin{equation}
    (e+p)Du^{\eta_s}+(\nabla^{\eta_s} + u^{\eta_s} D)p=F^{\eta_s\lambda}J_\lambda - u^{\eta_s}(e^\lambda J_\lambda).\label{inertia}
\end{equation}
Here, the derivative of $u$ and $p$ can be calculated as,
\begin{gather}
    Du^{\eta_s} = \frac{1}{\tau^2}\bar{\gamma}^2(\tau\partial_\tau + \bar{v}\partial_{\eta_s})Y,\label{derivative of u}\\
    (\nabla^{\eta_s} + u^{\eta_s} D)p=\frac{1}{\tau^2}\bar{\gamma}^2(\tau\bar{v}\partial_\tau + \partial_{\eta_s})p.\label{derivative of p}
\end{gather}
By substituting Eqs. (\ref{derivative of u}) and (\ref{derivative of p}) into Eq.~(\ref{inertia}), we obtain,
\begin{eqnarray}
    (e+&p)&(\tau\partial_\tau + \bar{v}\partial_{\eta_s})Y+(\tau\bar{v}\partial_\tau + \partial_{\eta_s})p\nonumber\\
    &=&\tau^2\bar{\gamma}^{-2}[F^{\eta_s\lambda}J_\lambda - u^{\eta_s} e^\lambda J_\lambda].\label{momentum}
\end{eqnarray}
The electric and magnetic four vectors are on the transverse plane, and they are perpendicular to each other,
\begin{gather}
    e^\mu = (0,e^x,0,0),\nonumber\\
    b^\mu = (0,0,b^y,0).\label{electromagnetic field in ALE}
\end{gather}
Then, Eqs.~(\ref{energy}) and (\ref{momentum}) are reduced to,
\begin{gather}
     (\tau\partial_\tau + \bar{v}\partial_\eta)e + (e+p)(\tau\bar{v}\partial_\tau Y + \partial_\eta Y)=\bar{\gamma}^{-1}\tau\sigma e_x^2,\label{fluid_ALE1}\\
      (e+p)(\tau\partial_\tau + \bar{v}\partial_{\eta_s})Y+(\tau\bar{v}\partial_\tau + \partial_{\eta_s})p=\bar{\gamma}^{-1}\tau\sigma e^x b^y.\label{fluid_ALE2}
\end{gather}
From the above assumption in Eqs.~(\ref{fluid velocity in ALE}) and (\ref{electromagnetic field in ALE}), Maxwell equations are written by,
\begin{gather}
    \partial_\tau\left[\left(u^\tau b^y+\frac{1}{\tau}e_x u_{\eta_s}\right)\right] + \partial_{\eta_s}\left[\left(u^{\eta_s}b^y-\frac{1}{\tau}e_x u_{\tau}\right)\right]\nonumber\\
    +\frac{1}{\tau}\left[\left(u^\tau b^y+\frac{1}{\tau}e_x u_{\eta_s}\right)\right]=0,\label{App_Maxwell_1}\\
    \partial_\tau\left[\left(u^\tau e^x+\frac{1}{\tau}b_y u_{\eta_s}\right)\right] + \partial_{\eta_s}\left[\left(u^{\eta_s}e^x-\frac{1}{\tau}b_y u_{\tau}\right)\right]\nonumber\\
    +\frac{1}{\tau}\left[\left(u^\tau e^x+\frac{1}{\tau}b_y u_{\eta_s}\right)\right]=-\sigma e_x.\label{App_Maxwell_2}
\end{gather}
In Ref.~\cite{PhysRevD.102.014017}, in order to solve these equations, we can take the following Ansatz:
\begin{gather}
    e_x(\tau,\eta_s) = -h(\tau,\eta_s)\sinh(Y-\eta_s),\\
    b_y(\tau,\eta_s) = h(\tau,\eta_s)\cosh(Y-\eta_s).
\end{gather}
Under this Ansatz, Eqs.~(\ref{App_Maxwell_1}) and (\ref{App_Maxwell_2}) give,
\begin{gather}
    \partial_\tau h(\tau,\eta_s) + \frac{h(\tau,\eta_s)}{\tau} = 0,\label{ALE_mm}\\
    \partial_{\eta_s}h(\tau,\eta_s)+\sigma\tau h(\tau,\eta_s)\sinh(\eta_s - Y) = 0,\label{ALE_mm2}
\end{gather}
and the solution of Eq.~(\ref{ALE_mm}) is found,
\begin{equation}
    h(\tau,\eta_s) = \frac{c(\eta_s)}{\tau},\label{ALE_h}
\end{equation}
where $c(\eta_s)$ is an arbitrary function.
By substituting Eq.~(\ref{ALE_h}) into Eq.~(\ref{ALE_mm2}), we obtain,
\begin{gather}
    \sinh(Y-\eta_s) = \frac{1}{\sigma\tau}\frac{\partial_{\eta_s}c(\eta_s)}{c(\eta_s)},\\
    \cosh(Y-\eta_s) = \sqrt{1+\left(\frac{1}{\sigma\tau}\frac{\partial_{\eta_s}c(\eta_s)}{c(\eta_s)}\right)^2}.
\end{gather}
Then, the fluid rapidity, four-velocity, and electromagnetic fields can be written as,
\begin{gather}
    Y = \eta_s + \sinh^{-1}\left(\frac{1}{\sigma\tau}\frac{\partial_{\eta_s}c(\eta_s)}{c(\eta_s)}\right),\\
    u^{\tau} = \sqrt{1+\left(\frac{1}{\sigma\tau}\frac{\partial_{\eta_s}c(\eta_s)}{c(\eta_s)}\right)^2},\\
    u^{\eta_s} = \frac{1}{\sigma\tau^2}\frac{\partial_{\eta_s}c(\eta_s)}{c(\eta_s)},\label{ALE_veta}\\
    e_x(\tau,\eta_s) = \frac{1}{\sigma\tau^2}\frac{\partial c(\eta_s)}{\partial\eta_s},\\
    b_y(\tau,\eta_s) = \frac{c(\eta_s)}{\tau}\times\sqrt{1+\left(\frac{1}{\sigma\tau}\frac{\partial_{\eta_s}c(\eta_s)}{c(\eta_s)}\right)^2}.
\end{gather}

Following to Ref.~\cite{PhysRevD.102.014017}, we take the form of the arbitrary function $c(\eta_s)$ as,
\begin{equation}
    c(\eta_s) = c_0\cosh(\alpha\eta_s),
\end{equation}
with the arbitrary parameters $\alpha$ and $c_0$.

In Ref.~\cite{PhysRevD.102.014017}, they split the conservation equation Eqs.~(\ref{fluid_ALE1}) and (\ref{fluid_ALE2}) into two ordinary differential equations (ODEs) with a given initial condition $e(\tau_0,0)=e_0$.
The combination of Eqs.~(\ref{fluid_ALE1}) and (\ref{fluid_ALE2}) with ideal gas EoS $p=\kappa e$ are obtained as,
\begin{gather}
    \partial_\tau e(\tau,\eta_s) + \frac{1+\kappa}{\tau}A(\tau,\eta_s)e(\tau,\eta_s) = B(\tau,\eta_s),\label{ce_ALE1}\\
    \partial_{\eta_s}e(\tau,\eta_s)+H(\tau,\eta_s)e(\tau,\eta_s) = G(\tau,\eta_s),\label{ce_ALE2}
\end{gather}
where $\kappa = 1/3$ and,
\begin{gather}
    A(\tau,\eta_s) = \left(\frac{\partial_{\eta_s}Y(\bar{v}-\kappa)-(\kappa-1)\tau\bar{v}\partial_\tau Y}{\kappa(\bar{v}^2-1)}\right),\\
    B(\tau,\eta_s) = \frac{\sigma(e_x b_y\bar{v}-\kappa e_x^2)}{\kappa\bar{\gamma}(\bar{v}^2-1)},\\
    H(\tau,\eta_s) = \frac{1}{\kappa}((1+\kappa)(\tau\partial_\tau Y+\bar{v}\partial_{\eta_s}Y))-(1+\kappa)\bar{v}A(\tau,\eta_s),\\
    G(\tau,\eta_s) = \frac{(\sigma\tau)e_x b_y}{\bar{\gamma}\kappa}-\tau\bar{v}B(\tau,\eta_s).
\end{gather}
We numerically solve Eqs.~(\ref{ce_ALE1}) and (\ref{ce_ALE2}) to obtain the profile and evolution of energy density.
First, we solve Eq.~(\ref{ce_ALE1}) as an ODE to find out the $\tau$-dependence of the function $e$, keeping constant as the variable $\eta_s$.
Then, we solve Eq.~(\ref{ce_ALE2}) with the solution of Eq.~(\ref{ce_ALE1}) in each $\tau$ as an initial condition of the ODE.
As a result, we get numerically the profile of the energy density as the solution of the ODEs.

\bibliography{Numerical_code_for_HIC}

\end{document}